\documentclass[prd,nofootinbib,twocolumn]{revtex4}

\usepackage{graphicx}
\usepackage{amssymb}
\usepackage{amsmath}
\usepackage{epstopdf}
\usepackage{aas_macros}
\usepackage{color}
\usepackage{hyperref}
\usepackage[dvipsnames]{xcolor}

\newcommand{\cpm}{{\small CUBEP$^3$M}}
\newcommand{\class}{{\small CLASS}}

\newcommand{\kms}{{\rm km}/{\rm s}}
\newcommand{\bh}{{\rm PBH}}
\newcommand{\dm}{{\rm PDM}}
\newcommand{\hl}{{\rm HL}}
\newcommand{\ucmh}{{\rm UCMH}}
\newcommand{\lcdm}{{$\Lambda$CDM}} 
\newcommand{\lpdm}{{$\Lambda$PDM}} 
\newcommand{\lpbh}{$\Lambda$PBH}
\newcommand{\zeq}{z_{\rm eq}}

\begin{document}

\title{Early Structure Formation in \lpbh~Cosmologies}

\author{Derek Inman} \email{derek.inman@nyu.edu} \affiliation{Center
  for Cosmology and Particle Physics, Department of Physics, New York
  University, 726 Broadway, New York, NY, 10003, USA}

\author{Yacine Ali-Ha\"{i}moud} \email{yah2@nyu.edu}
\affiliation{Center for Cosmology and Particle Physics, Department of
  Physics, New York University, 726 Broadway, New York, NY, 10003,
  USA}

\begin{abstract}

  Cold dark matter (CDM) could be composed of primordial black holes
  (\bh) in addition to or instead of more orthodox weakly interacting
  massive particle dark matter (\dm).  We study the formation of the first structures in such \lpbh{} cosmologies using $N$-body simulations
  evolved from deep in the radiation era to redshift 99.  When \bh{}
  are only a small component of the CDM, they are clothed by \dm{} to
  form isolated halos.  On the other hand, when \bh{} make most of the
  CDM, halos can also grow via clustering of many \bh.  We find that
  the halo mass function is well modelled via Poisson statistics
  assuming random initial conditions.  We quantify the nonlinear
  velocities induced by structure formation and find that they are too
  small to significantly impact CMB constraints. A chief challenge is
  how best to extrapolate our results to lower redshifts relevant for
  some observational constraints.
  
\end{abstract}

\maketitle

\begin{section}{Introduction}

  Despite overwhelming evidence for cold dark matter (CDM), we have no
  knowledge of what it is composed of.  The prevailing candidate has
  been a Weakly Interacting Massive Particle (WIMP) which freezes out
  in the early Universe as a cold relic \citep{bib:Jungman1996}.
  However, there has been neither direct nor indirect detection of
  such a WIMP.  Furthermore, the predictions of a completely cold
  species is in potential tension with some observations of small-scale structure \citep{bib:Bullock2017}.  Alternative particle
  explanations of CDM include warm \citep{bib:Dodelson1994}, fuzzy
  \citep{bib:Hu2000} and axion dark matter \citep{bib:Marsh2016}, none
  of which have been detected.

  It is also possible to consider non-particle based candidates for
  CDM, of which primordial black holes (\bh) are a long-studied
  favorite \citep{bib:Zeldovich1967,bib:Hawking1971,bib:Carr1974}
  (see, e.g., \citep{bib:Carr2016,bib:GarciaBellido2017} for more
  modern discussions).  Detections by LIGO of the merger of black hole
  binaries with masses of order $\sim10~M_\odot$
  \citep{bib:Ligo2016a,bib:Ligo2016b,bib:Ligo2017a,bib:Ligo2017b,bib:Ligo2017c}
  have led to the suggestion that the coalescing black holes may have
  been primordial and could make up (at least part of) the CDM
  \citep{bib:Bird2016, bib:Clesse2017, bib:Sasaki2016}.  Even if PBH
  do not make all of the dark matter, the possibility of both
  WIMP-like particles (which we will generically call \dm{} for
  ``Particle Dark Matter'') and \bh{} coexisting and interacting
  gravitationally will have novel phenomenology.  In this paper, we
  perform numerical calculations of mixed PBH-PDM dark matter, and
  examine their nonlinear dynamics down to redshift 99.

  In addition to being a bona fide dark matter candidate, PBH can
  provide a window onto the initial conditions in the Universe.  On
  large scales, the initial power spectrum has been precisely
  constrained to be nearly scale invariant with overdensities of
  $\lesssim10^{-4}$ \citep{bib:Planck2015}.  However, the power
  spectrum on ultra small scales is only poorly constrained (see,
  e.g., \citep{bib:Jeong2014,bib:Nakama2014}), and could be
  significantly larger. Enhanced initial overdensities on small scales
  can yield very rich phenomena.  If they are greater than
  $\gtrsim 10^{-3}$ then standard adiabatic \dm{} perturbations form
  bound nonlinear structures almost immediately at matter radiation
  equality \citep{bib:Ricotti2009}.  Such structures were thought to
  be extremely dense, and so called ultracompact minihalos (\ucmh),
  but recent numerical work has indicated that this is not the case
  due to frequent mergers
  \citep{bib:Gosenca2017,bib:StenDelos2018,bib:StenDelos2018b}.  An
  alternative way to produce early structure formation is a strong
  blue-tilt to the power spectrum \citep{bib:Hirano2015}.  Larger
  perturbations can form \dm{} ``clumps" in radiation domination
  \citep{bib:Berezinsky2010} and even cause shocks in the radiation
  fluid \citep{bib:Pen2016}.  Increasing the primordial amplitude even
  further, $\gtrsim10^{-1}$, will cause \bh{} to form
  \citep{bib:Green2004}.  Such \bh{} form {\it regardless} of whether
  there is some alternate form of \dm, and so are their own unique CDM
  candidate.

  On sufficiently large scales, we expect the PBH density field to
  follow the standard adiabatic perturbations. However, on small
  enough scales, the discrete nature of PBH becomes important. If
  \bh{} make up only a small fraction of the CDM then we expect them
  to be clothed by a large amount of \dm{}, but not interact with any
  other \bh.  This allows for analytic treatments of accretion, e.g.,
  \citep{bib:Ricotti2007,bib:Mack2007,bib:Ricotti2009,bib:Rice2017,bib:Lacki2010,bib:Eroshenko2016},
  and the resulting halos (\hl) are expected to be similar to the
  theorized \ucmh{} with a very steep density profile
  \citep{bib:Bertschinger1985}.  In the opposing limit, when \bh{} are
  the entirety of the CDM, the large-scale behaviour is still the
  adiabatic growing mode, but on small scales the random locations of
  \bh{} introduces a shot noise in their density
  \citep{bib:Afshordi2003}.  These two limits have respectively been
  called the ``seed'' and ``Poisson'' limits and have ramifications
  for supermassive black hole formation \citep{bib:Carr2018}.
  Intermediate cases, on the other hand, are difficult to study
  analytically due to the competition of \bh{} in accreting \dm{} and
  \bh-\bh{} interactions.

  It is important to understand these highly nonlinear dynamics in
  order to make accurate constraints on the amount of \bh.  For
  instance, gravitational constraints which assume a uniform density,
  such as those from microlensing \citep{bib:Alcock2000} and dynamical
  heating \citep{bib:Brandt2016} could be affected if \bh{} become
  highly clustered at late times (e.g.~\cite{Metcalf_96}).  The fact
  that \bh{} could have a steep \dm{} profile could affect potential
  constraints coming from pulsar timing \citep{bib:Schutz2017} where
  the associated \dm{} halo could also induce Shapiro delay
  \citep{bib:Clark2016,bib:Clark2017}.  Such a steep \dm{} profile
  could also cause \dm{} annihilation to be observed
  \citep{bib:Lacki2010}, although this assumes a specific form for the
  \dm.  Furthermore, Poisson fluctuations can cause \bh{} binaries to
  form in the early Universe \citep{bib:Nakamura1997} which could then
  merge and be detected by LIGO \citep{bib:AliHaimoud2017b,
    bib:Raidal2017}.  While analytic estimates suggest that the tidal
  field on these binaries is insufficient to disrupt them
  \cite{bib:Ioka1998, bib:AliHaimoud2017b}, it is possible that
  nonlinear effects could affect this conclusion
  \cite{bib:Raidal2018}.  Such constraints come from astrophysical
  observations at late times in the local Universe.  The CMB also
  provides constraints on \bh{} abundances \cite{bib:Miller2000,
    bib:Miller2001, Ricotti_08}, and is sensitive to relative motion
  between \bh{} and gas in the Universe which will include a nonlinear
  component \citep{bib:AliHaimoud2017a,bib:Poulin2017}.

  \begin{figure*}
    \includegraphics[width=0.85\textwidth]{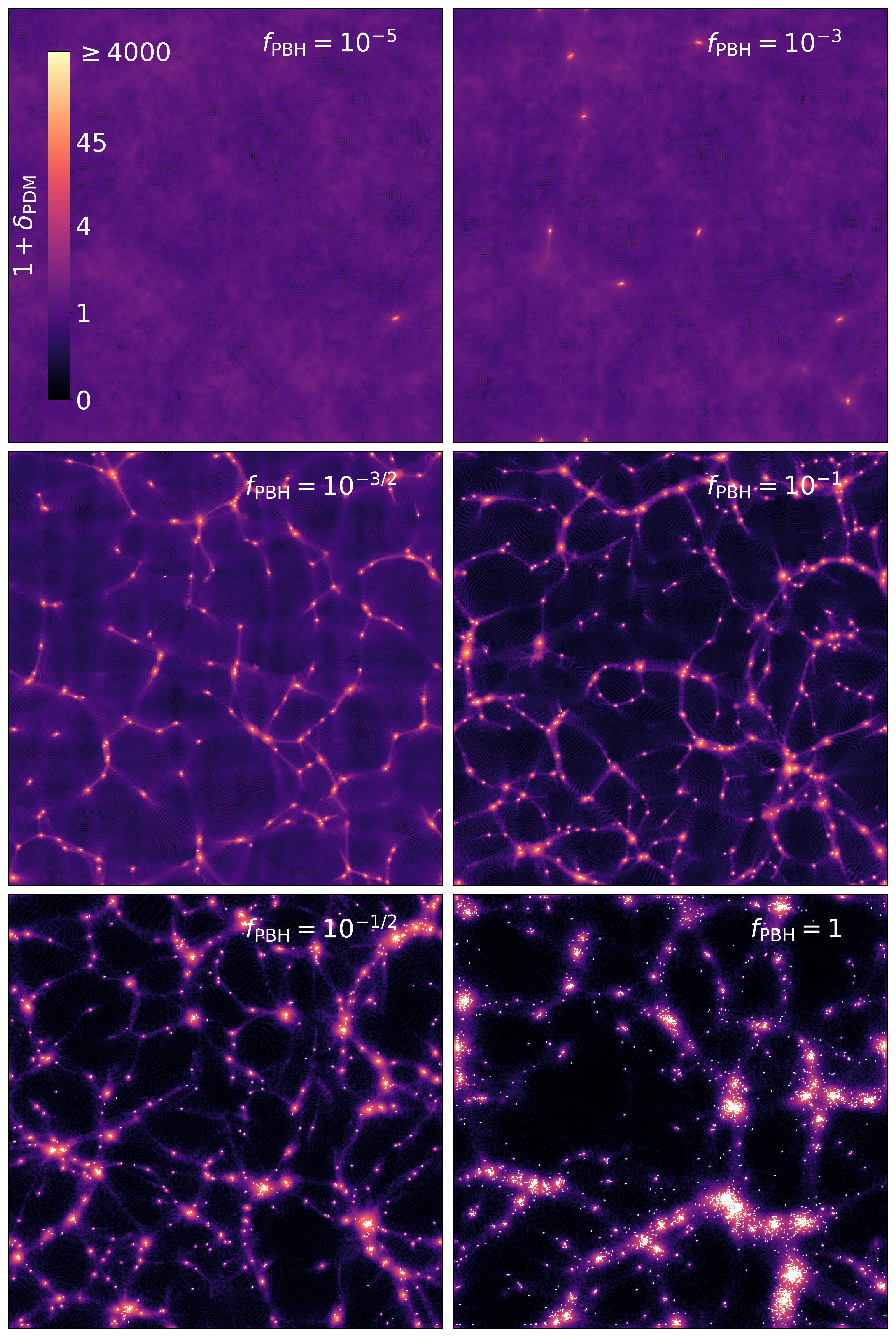}
    \caption{The matter field at $z=99$ for various \bh{} fractions.
      The \dm{} density field is represented by the colormap with
      white points indicating \bh{} locations.  The slice width is
      1/16 the box size, just under 2 kpc/$h$.}
    \label{fig:slices}
  \end{figure*}

  To correctly understand such nonlinear dynamics we must utilize
  numerical simulations.  In this work, we develop $N$-body simulations
  which evolve both \dm{} and \bh{} particles and analyze the
  resulting halo characteristics.  We focus on the scenario where
  $M_\bh= 20~h^{-1} M_\odot$ but vary the relative fraction of \bh{}
  to \dm.  We illustrate our results in Fig. \ref{fig:slices} which
  shows the \dm{} density fields alongside points to indicate where
  \bh{} and halos are.  The differences between ``\lpdm{}'' (top left
  panel) and ``\lpbh{}'' (bottom right panel) are quite dramatic!  In
  the next section we describe how these simulations were performed.
  We then show our results for \dm{} clustering around \bh{}, the halo
  mass function, and the distribution of \bh{} velocities, which we
  argue do not significantly affect CMB constraints. Lastly, we
  conclude by discussing how these simulations can next be used to
  improve constraints based on primordial \bh{} binaries and also how
  they may affect the formation of first stars at cosmic dawn.
  
\end{section}

\begin{section}{Cosmology}
  \label{sec:cosmo}
  \begin{subsection}{Background cosmology}

    Before describing the computations in more detail, it is worth
    going through the cosmic inventory to give an overview of how we
    treat different components and values of cosmological parameters
    used in the simulations. Throughout this work, we use cosmological
    parameters consistent with \emph{Planck} results
    \citep{bib:Planck2015}.

    $\bullet$ {\it Matter -- } The matter sector consists of \bh{}
    (which for simplicity we assume to all have the same mass
    $M_{\bh}$), \dm{} and baryons and contributes
    $\Omega_m = \Omega_c + \Omega_b$ to the energy density. The \bh{}
    and \dm{} contribute $\Omega_c=0.26$ to the energy density and
    their individual contributions are parameterized via $f_{\bh}$
    such that $\Omega_{\bh} =f_{\bh}\Omega_c$ and
    $\Omega_{\dm}=(1-f_{\bh})\Omega_c$. The baryons contribute
    $\Omega_b=0.05$.  We do not include hydrodynamics in our
    simulations and therefore treat the baryons as an unclustered
    species, adding a constant $\Omega_b/\Omega_m$ to the grid.  This
    is a very good approximation at early times when baryons are
    coupled to photons via Compton scattering, but becomes
    progressively worse as they begin to catch up to the \dm{} and
    cluster around \bh. For the purely gravitational effects of
    primary interest in this work, the neglect of baryon clustering
    leads to errors of order $\Omega_b/\Omega_m$.

    $\bullet$ {\it Radiation -- } Neutrinos and photons, collectively
    radiation, dominate the energy budget in the early stages of the
    Universe.  Their energy density is parameterized by the redshift
    of matter-radiation equality, $\zeq=3374$, such that
    $\Omega_r=\Omega_m/(1+\zeq)$. Photons and neutrinos are
    essentially free-streaming on the scales relevant to our
    simulations, so we assume they are homogeneous and only contribute
    to the expansion rate.  We assume the neutrinos are massless,
    although this should have little effect on our results as they are
    largely relativistic at the times of interest.

    $\bullet$ {\it Dark Energy and Curvature -- } Dark energy takes up
    whatever is needed to have a flat, critical density Universe:
    $\Omega_\Lambda = 1 - \Omega_r - \Omega_m$.  At the redshifts
    considered here, the dynamical effects of dark energy are
    completely negligible, although for definiteness we assume a
    standard cosmological constant.

    The expansion of the Universe is therefore given by the standard
    Hubble equation:
    \begin{align}
      H= \frac{d \ln a}{dt} = H_0\sqrt{\Omega_\Lambda + \Omega_m a^{-3} + \Omega_r a^{-4}}.
      \label{eq:hubble}
    \end{align}
    with $H_0 = 100~h~ {\rm km/s/Mpc}$ with $h=0.67$, although the
    simulations are only sensitive to this value through the initial
    conditions.

  \end{subsection}
   
  \begin{subsection}{Initial perturbations}

    The most natural way to form PBHs is through enhanced primordial
    curvature perturbations, collapsing into black holes upon horizon
    entry \cite{bib:Carr1975}. The PBH mass depends on the initial
    overdensity, but is typically comparable to the mass-energy inside
    the Hubble radius at the time of collapse
    \cite{bib:Niemeyer1998}. The relationship between the scale $k_*$
    at which the power spectrum is enhanced and the characteristic PBH
    mass is then \cite{Sasaki_18}
    \begin{align}
      M_\bh \approx 30 ~ M_{\odot} \left( \frac{300 ~
      \textrm{kpc}^{-1}}{k_*}\right)^2.  
    \end{align}
    For comparison, the comoving scales relevant to our simulations
    are $k \approx 0.15 - 38$ kpc$^{-1}$, as we will discuss in the
    next section. The smallest modes simulated are therefore only
    separated from the PBH scale by a factor of 8 or so.
    
    For Gaussian initial conditions, the variance of curvature
    perturbations required to form PBHs is $\Delta^2 \approx 0.03$
    (see e.g.~\cite{bib:Germani2019, bib:Young2019} and references
    therein). The abundance of PBHs typically depends exponentially on
    $\Delta^2$, and conversely, the detailed amplitude only depends
    logarithmically on $f_\bh$. It is therefore likely that the
    initial curvature perturbation is still rather large on the
    smallest scales we simulate. Ref.~\cite{Byrnes_18} showed that,
    for primordial perturbations generated during single-field
    inflation, the steepest possible growth of the primordial
    curvature power spectrum (per $\ln k$) scales as $k^4$. Therefore,
    within single-field inflation, and neglecting non-Gaussianities,
    the existence of PBHs would require $\Delta^2 \gtrsim 10^{-5}$ on
    the simulation grid scale, significantly larger than the amplitude
    on scales probed by CMB-anisotropy measurements, with
    $\Delta^2_{\rm CMB} \approx 2 \times 10^{-9}$. On scales
    $k \lesssim 5$ kpc$^{-1}$, $\Delta^2 \approx \Delta_{\rm CMB}^2$
    is compatible with PBH formation (at least within the
    single-field, Gaussian approximation). Also note that upper limits
    on CMB spectral distortions imply that \cite{Chluba_12}
    $\Delta^2 \lesssim 3 \times 10^{-5}$ for
    $0.1 \lesssim k~ \textrm{kpc} \lesssim 5$.
    
    For simplicity, we assume a primordial curvature power spectrum
    extrapolated from CMB scales, with a constant spectral index,
    i.e.~
    \begin{align}
      \Delta^2(k) &= 2.215 \times 10^{-9} \left(\frac{k}{0.075~h/\textrm{kpc}}\right)^{-0.038} \nonumber\\
                  &= 1.544 \times 10^{-9}
                    \left(\frac{k}{1~h/\textrm{kpc}}\right)^{-0.038}. \label{eq:Delta-ad}
    \end{align}
    As we discussed above, on the smallest scales of our simulations,
    this is inconsistent with PBH formation in the case of
    single-field-inflation with Gaussian initial conditions. It may still
    be consistent with PBH formation for different sets of
    assumptions. Moreover, we do not expect this choice to affect any
    of our results; indeed, the initial adiabatic perturbations of PDM
    are quickly overwhelmed by isocurvature perturbations due to PBHs,
    as we will see later on.

    We assume that the primordial PDM perturbations are adiabatic, as
    are PBH fluctuations on sufficiently large scales. On small enough
    scales, however, we assume PBHs are effectively randomly
    distributed \cite{bib:AliHaimoud2018, Ballesteros_18,
      Desjacques_18}. The variance of initial Poisson perturbations
    (per $\ln k$) takes the form
    \begin{align}
      \Delta_\bh^2(k) = (k/k_*)^3,
    \end{align}
    where $k_*$ is of order the inverse mean separation between PBHs:
    \begin{align}
      k_* = \left(2 \pi^2 \overline{n}_{\rm bh}\right)^{1/3}
      \approx 4 ~ h/\textrm{kpc}~ f_\bh^{1/3} \left(\frac{20 ~
      M_{\odot}/h}{M_\bh}\right)^{1/3}.  
    \end{align}
    On the $\sim$ kpc scales of our simulations, discreteness noise
    largely overwhelms the intrinsic adiabatic perturbations of PBHs.

  \end{subsection}

\end{section}

\begin{section}{Methods}

  The $N$-body code used here is a modified version of \cpm{}, a fast,
  massively parallel cosmological simulation code
  \citep{bib:Merz2005,bib:HarnoisDeraps2013,bib:Emberson2017}.  The
  addition of \bh{} builds off the neutrino integration described in
  \citep{bib:Inman2015,bib:Emberson2017}.

  \begin{subsection}{Simulation setup}
    \label{sec:setup}

    For a given fraction $f_\bh$ of CDM in \bh{} of mass $M_{\bh}$, we
    need to select the optimal numerical parameters for our
    simulations. These parameters are the number of $N$-body particles
    used to evolve the \bh{} and \dm{} ($N_{\bh}$ and $N_\dm$,
    respectively) and the simulation box size $L$. Given the box size
    and number of particles, the $N$-body particle mass is then given
    by
    \begin{align}
      m_i = 278~ h^{-1} M_{\odot} \left(h^{-1}\textrm{kpc}\right)^{-3}\Omega_i
      (L^3/N_i). 
      \label{eq:m-to-N} 
    \end{align}
    We are interested in studying the effects arising from the
    \emph{discreteness} of PBH, and therefore set the PBH particle
    mass to the \emph{physical} \bh{} mass, $m_\bh = M_\bh$. In
    contrast, it is virtually impossible to simulate individual \dm{}
    particles due to their large abundance; in that case, $N$-body
    particles represent, as is standard, chunks of phase-space. To
    maximize accuracy, we assign as large a number of \dm{} particles
    as allowed by computational resources. Our fiducial setup has
    $N_\dm = N_{\rm fid} \equiv 2 \times 256^3$.  These considerations
    determine two combinations of the three parameters
    $(N_\bh, N_\dm, L)$. To determine the third, we enforce that the
    \emph{artificial} Poisson noise from PDM $N$-body particles is
    subdominant to the \emph{physical} Poisson fluctuations in the PBH
    abundance. Poisson noise is inversely proportional to the number
    of particles (for a fixed box size), so this criterion amounts to
    \begin{align}
      \frac{(1-f_\bh)^2}{N_\dm} \ll \frac{f_\bh^2}{N_\bh},
    \end{align}
    where the factors $(1-f_\bh)$ and $f_\bh$ weigh the respective
    contributions to the total mass density perturbation. Using
    Eq.~\eqref{eq:m-to-N} with $m_{\bh} = M_\bh$, we rewrite this as
    \begin{align}
      \left(\frac{L}{h^{-1} \textrm{kpc}}\right)^3 \ll
      \frac{M_{\bh}}{72 ~h^{-1} M_\odot} N_{\bh} \frac{f_\bh}{(1 -
      f_\bh)^2}.  
    \end{align}
    This limits the box size to
    \begin{align}
      L &\lesssim 200~h^{-1} \textrm{kpc}\left(\frac{M_{\bh}}{20 ~h^{-1} M_\odot} \frac{N_{\dm}}{N_{\rm fid}} \right)^{1/3}\nonumber\\
        &\times \frac{f_\bh^{1/3}}{(1 - f_\bh)^{2/3}}. 
          \label{eq:max-L}
    \end{align}
    Rather than adjusting $L$ for each value of $f_\bh$, we choose to
    have a single box size, to facilitate comparisons between
    simulations. We pick $L = 30 ~h^{-1}$ kpc for
    $M_\bh = 20~h^{-1} M_{\odot}$, which satisfies the criterion
    \eqref{eq:max-L} for $f_\bh \gtrsim 10^{-3}$.  In simulations with
    lower $f_\bh$, discreteness noise from PDM particles is less of an
    issue. Instead, we want the \dm{} halos that form around any \bh{}
    to be well resolved by many particles once the \dm{} halo mass
    becomes comparable to the \bh{} mass.  This then implies that the
    mass ratio
    \begin{align}
      \frac{m_\bh}{m_\dm} = \frac{f_\bh}{1-f_\bh} \frac{N_\dm}{N_\bh}
    \end{align}
    must be kept large to properly resolve the halo formation.  As an
    example, our simulation with $f_\bh=10^{-5}$ containing just a
    single \bh{} has a halo containing thousands of \dm{} particles,
    whereas the ratio of $N_\dm/N_\bh$ is only a few hundred for the
    simulation with $f_\bh=1$.  In practice, this limits the value of
    $f_\bh$ to
    \begin{align}
      f_\bh \gg \frac{1}{1+\frac{N_\dm}{N_\bh}}
    \end{align}
    which for our choice of parameters implies
    $f_\bh \gg N_\bh/N_\dm \ge 3\times10^{-8}$.
    
    With these parameters, the number of PBH in the simulation box is
    $N_\bh = 10^5 f_\bh$, and the mass of the \dm{} $N$-body particles
    is $m_{\dm} \approx 0.058~ h^{-1} M_{\odot} (1 - f_\bh)$. For
    simulations with $f_\bh=1$ where $m_\dm=0$, we opt to still evolve
    the \dm{} particles as \emph{tracer} particles, hence neglect
    \dm{}-\dm{} forces. We keep PDM particles in our
    $f_\bh \rightarrow 1$ simulations as even a small fraction of PDM
    could have noticeable effects (for instance, if the PDM self
    annihilates \citep{bib:Lacki2010,bib:Adamek2019}.
    
  \end{subsection}

  \begin{subsection}{Initial Conditions} \label{ssec:ic}
    
    \begin{subsubsection}{Large scales} \label{sec:pert-theory}
    
      In this section we focus on scales larger than the
      characteristic inter-PBH separation,
      $k \lesssim \overline{n}_{\bh}^{1/3}$, so that we can treat PBHs
      as a quasi-homogeneous ideal pressureless fluid.
    
      We assume that the \emph{primordial} PDM perturbations
      (i.e.~prior to horizon entry, and labelled by the superscript 0)
      are purely adiabatic on all scales,
      $\delta_{\dm}^0 = \delta_{\rm ad}^0$ and
      $\dot{\delta}_{\dm}^0 = \dot{\delta}_{\rm ad}^0$. We decompose
      the primordial PBH overdensity into an adiabatic piece and an
      uncorrelated isocurvature piece:
      $\delta_\bh^0 \equiv \delta_{\rm ad}^0 + \delta_{\rm iso}^0$.
      The latter is due to the discreteness of PBHs, which prevents
      them from being distributed strictly adiabatically on small
      scales. It largely dominates over the adiabatic piece over all
      scales of our simulation. We assume that PBHs are formed with
      negligible velocities relative to the PDM, so that
      $\dot{\delta}_{\bh}^0 = \dot{\delta}_{\dm}^0 = \dot{\delta}_{\rm
        ad}^0$.

      We define the total dark matter overdensity as
      $\delta_c \equiv (1 - f_\bh) \delta_\dm + f_\bh \delta_\bh$, and
      the difference field as
      $\delta_- \equiv \delta_\bh - \delta_\dm$. The PDM and PBH
      density fields are related to those through
      \begin{align}
        \delta_\dm &= \delta_c - f_\bh \delta_-, \label{eq:c-to-pdm}\\
        \delta_\bh &= \delta_c + (1 - f_\bh)
                     \delta_-.
                     \label{eq:c-to-pbh}
      \end{align} 
      The initial conditions for the total CDM perturbation are
      $\delta_c^0 = \delta_{\rm ad}^0 + f_\bh \delta_{\rm iso}^0$,
      $\dot{\delta}_c^0 = \dot{\delta}_{\rm ad}^0$. We denote by
      $T_{\rm ad}(a)$ and $T_{\rm iso}(a)$ the linear transfer
      functions of the adiabatic and CDM density isocurvature modes,
      respectively, both normalized to unity at $a \rightarrow 0$. The
      CDM overdensity at scale factor $a$ is then
      \begin{align}
        \delta_c(a) =
        T_{\rm ad}(a) \delta_{\rm ad}^0 + T_{\rm iso}(a) f_\bh
        \delta_{\rm iso}^0. \label{eq:delta_c(a)}
      \end{align}
      Since both PDM and PBHs are cold fluids, they are subject to the
      same forces, and their (gauge-invariant) relative velocity
      decays as $1/a$. Having assumed a negligible primordial relative
      velocity, we conclude that it remains so at all times. As a
      consequence, the difference of the linearized continuity
      equations imply that
      $\delta_- = \delta_-^0 = \delta_{\rm iso}^0$ is
      constant. Inserting Eq.~\eqref{eq:delta_c(a)} into
      Eqs.~\eqref{eq:c-to-pdm} and \eqref{eq:c-to-pbh}, we obtain
      \begin{align}
        \delta_\dm(a) &= T_{\rm ad}(a) \delta_{\rm ad}^0 + \left(T_{\rm iso}(a) -1 \right) f_\bh \delta_{\rm iso}^0, \label{eq:delta_PDM-lin}\\
        \delta_\bh(a) &= \delta_{\rm iso}^0 + \delta_{\dm}(a).
                        \label{eq:delta_PBH-lin} 
      \end{align}
      Both adiabatic and isocurvature transfer functions can be
      extracted from a Boltzmann code, but it is relatively
      straightforward to explicitly compute $T_{\rm iso}$
      analytically. Let us first focus on times well after horizon
      entry (or equivalently, on deeply sub-horizon scales). Radiation
      (photon-baryons and neutrinos) perturbations fluctuate on rapid
      timescales, and the evolution of the \emph{slow} mode of total
      CDM perturbations is given by \cite{Hu_96, Weinberg_02,
        Voruz_14}
      \begin{align}
        \delta_c'' + \frac{3 s + 2}{2 s(s+1)}\delta_c'
        - \frac{3 \gamma}{2 s(s+1)} \delta_c= 0, 
      \end{align}
      where $\gamma \equiv \Omega_c/\Omega_m$ is the fraction of
      matter that clusters (baryons being unclustered), and primes
      denote differentiation with respect to $s = a/a_{\rm eq}$. This
      equation has two independent solutions, which generalize the
      Meszaros solutions \cite{Meszaros_74} to $\gamma < 1$.
      Ref.~\cite{Hu_96} gives explicit expressions in terms of
      hypergeometric functions; however, both of their solutions
      diverge logarithmically at $s \rightarrow 0$ for $\gamma < 1$.
      Instead, we define our two independent solutions
      $D_{+}(s), D_-(s)$ as follows:
      \begin{align}
        D_+(s) &= ~_2F_1\left(- \alpha_-, \alpha_+, 1, -s\right), \\
        D_-(s) &= (1 + s)^{- \alpha_+} \nonumber\\
               & \times ~ _2F_1\left(\alpha_+, \alpha_+ + \frac12, 2 \alpha_+ + \frac12, \frac1{1 + s}\right),\\
        \alpha_{\pm} &\equiv \frac14 \left(\sqrt{1 + 24 \gamma} \pm 1
                       \right).  
      \end{align}
      The solution $D_-$ diverges logarithmically as
      $s \rightarrow 0$, but $D_+ \rightarrow 1$ for
      $s \rightarrow 0$, for any $\gamma$.

      The adiabatic and isocurvature transfer functions are linear
      combinations of $D_+, D_-$. The coefficients are found by
      matching at $s \rightarrow 0$ with the asymptotic limits of the
      solutions to the exact relativistic equations (this applies for
      modes entering the horizon well inside radiation
      domination). For the adiabatic mode, this gives the well-known
      logarithmic growth during the radiation era, since matching
      requires a non-negligible contribution from $D_-$
      \cite{Weinberg_02, Voruz_14}. For the isocurvature mode,
      however, CDM perturbations remain constant through horizon
      crossing and during radiation domination (see
      e.g.~Ref.~\cite{bib:Bucher2000} and the appendix of
      Ref.~\cite{bib:Chluba2013}). Therefore,
      $T_{\rm iso}(a) = D_+(a/a_{\rm eq})$. The asymptotic behaviours
      of $D_+$ are
      \begin{align}
        D_+(s) &\approx 1 + \frac{3 \gamma}{2}s , \ \ s  \ll 1, \\
        D_+(s) &\propto s^{\alpha_-}, \ \ s \gg 1.  
      \end{align}
      The following very simple analytic expression fits the exact
      isocurvature transfer function to better than $1.5\%$ accuracy
      for $0.5 \leq \gamma \leq 1$ and for all values of $s$:
      \begin{align}
        D_+(s)
        \approx \left(1 + \frac{3 \gamma}{2 \alpha_-}
        s\right)^{\alpha_-}. \label{eq:D+-approx} 
      \end{align}
      Note that this simple fit is very accurate due to a coincidence:
      the coefficient of $s^{\alpha_-}$ at large $s$ happens to be
      close to $(3 \gamma/2 \alpha_-)^{\alpha_-}$ for
      $\gamma \approx 1$.

      In summary, deep in the radiation era, and for scales larger
      than the characteristic inter-PBH separation, we have shown that
      \begin{align}
        \delta_{\dm}(a) &\approx T_{\rm ad}(a) \delta_{\rm ad}^0 +
                          \frac32 \gamma \frac{a}{a_{\rm eq}}
                          f_\bh \delta_{\bh}^0,  
                          \label{eq:delta_dm(a)} \\
        \delta_{\bh}(a) &\approx \delta_{\bh}^0 + \delta_{\dm}(a), 
                          \label{eq:delta_pbh(a)} 
      \end{align}
      where we have replaced
      $\delta_{\rm iso}^0 \approx \delta_\bh^0$, and neglected the
      term $a/a_{\rm eq} f_\bh \delta_{\rm iso}^0$ in the PBH density
      field, as it is always small relative to $\delta_{\rm iso}^0$.

      From these linear density fields and their derivatives, one can
      obtain the displacement fields $\vec{\psi}$ in the Zeldovich
      approximation \citep{bib:Zeldovich1970}, through
      $\vec{\nabla} \cdot \vec{\psi} = - \delta$, or, in Fourier
      space, $\vec{\psi}(\vec{k}) = i \hat{k} \delta(\vec{k})/k$. The
      velocity fields $\dot{\vec{\psi}}$ are obtained from the time
      derivative of these equations.

    \end{subsubsection}

    \begin{subsubsection}{Small scales} \label{sec:small-scales}

      On scales smaller than the inter-PBH separation, the PBH density
      field is formally non-linear from the get-go, and we can no
      longer use linear perturbation theory. For simplicity, and for
      lack of a better theory, we assume that the adiabatic
      contributions of Eqs.~\eqref{eq:delta_dm(a)} and
      \eqref{eq:delta_pbh(a)} hold down to arbitrarily small
      scales. Since the adiabatic initial perturbation is nearly
      scale-invariant, and the transfer function only induces a
      logarithmic dependence on $k$, the variance of the adiabatic
      displacement field (per $\ln k$) scales as
      $\langle \psi_{\rm ad}^2 \rangle \sim 1/k^2$, and is dominated
      by large scales, so this approximation should not lead to
      substantial errors.

      To obtain the isocurvature displacements for the PBH and PDM, we
      solve Hamilton's equations \citep{bib:Bertschinger1993} for
      particles around PBHs.  We denote by $\vec{x}$ the initial,
      unperturbed comoving position, and by $\vec{\psi}(\vec{x})$ the
      displacement field. Hamilton's equations read
      \begin{align}
        \frac{d}{d
        \tau} \left(a \frac{d \vec{\psi}}{d \tau} \right) =
        -a\vec{\nabla}\phi(\tau, \vec{x} +
        \vec{\psi}). \label{eq:Hamilton} 
      \end{align}
      The potential $\phi$ is comprised of two pieces. First, the PBH
      contribution $\phi_{\rm iso}$, given by
      \begin{align}
        a\vec{\nabla}\phi_{\rm iso}(\tau,\vec{x}+\vec{\psi})  &= -4 \pi G M_\bh \sum_i \frac{\vec{x} - \vec{x}_i}{|\vec{x} - \vec{x}_i|^3}, \\
        a \nabla^2\phi_{\rm iso}(\tau,\vec{x}+\vec{\psi}) &= 4 \pi G M_\bh \sum_i \delta_{\rm D}(\vec{x} - \vec{x}_i) \nonumber\\
                                                              &= \frac32 \gamma f_\bh H_0^2 \Omega_m \delta_\bh
                                                                \label{eq:phi_iso}, 
      \end{align}
      where $\vec{x}_i=\vec{x}_i(\tau)$ are the PBH positions. In
      addition, $\phi$ gets a contribution from the PDM as it clusters
      around PBHs.
      
      We solve Hamilton's equations by making the Born approximation
      for both PDM and PBH,
      i.e.~$\vec{x} + \vec{\psi} \approx \vec{x}$ in the right-hand
      side of Eq.~\eqref{eq:Hamilton}.  This means that we neglect any
      induced PDM overdensities, $\phi=\phi_{\rm iso}$, and that we
      take PBH to be stationary so that $\delta_\bh=\delta_\bh^0$.  We
      therefore solve
      \begin{align}
        \label{eq:Hamilton2}
        \frac{d}{d \tau} \left(a \frac{d \vec{\psi}}{d \tau} \right) &=
                                                                       -a\vec{\nabla}\phi_{\rm
                                                                       iso}(\vec{x}).  
      \end{align}
      The solution is given by
      \begin{align}
        \vec{\psi} &= - \frac{4}{\Omega_mH_0^2} \mathcal{F}(s) a \vec{\nabla} \phi_{\rm iso}, \\
        \mathcal{F}(s) &\equiv
                         \log\left[\frac{1}{2}\left(1+\sqrt{1+s}\right)\right]. \label{eq:icfactors}
      \end{align}
      The Born approximation only holds as long as
      $|\vec{\psi}| \ll |\vec{x}- \vec{x}_i|$. Replacing
      $\phi_{\rm iso}$ by its explicit expression, this condition can
      be rewritten as
      \begin{align}
        |\vec{x}- \vec{x}_i|^3 \gg V \equiv f_\bh \overline{n}_\bh^{-1} \mathcal{F}(s). \label{eq:Born-validity}
      \end{align}
      The PDM contained within a PBH region of influence $V$ is
      effectively bound to it; the total mass of PDM bound to a PBH is
      therefore approximately
      \begin{align}
        M_{\rm bound} &\sim (1 - f_\bh)\overline{\rho}_c V \nonumber\\
                      &\sim (1 - f_\bh) M_\bh \mathcal{F}(s).  
      \end{align} 
      Our neglect of the potential sourced by the induced PDM
      overdensity requires $M_{\rm bound} \ll M_\bh$, hence
      \begin{align}
        (1 -
        f_\bh)\mathcal{F}(s) \ll 1.  
      \end{align}
      If we moreover require the Born approximation to hold for scales
      comparable to the inter-PBH separation and smaller,
      Eq.~\eqref{eq:Born-validity} implies
      \begin{align}
        f_\bh \mathcal{F}(s) \ll 1.
      \end{align}
      For both approximations to be valid, we therefore require
      $\mathcal{F}(s) \ll 1$, implying $s \ll 1$,
      i.e.~$a \ll a_{\rm eq}$.  In this limit, we hence obtain the PDM
      displacement
      \begin{align}
        \vec{\psi} \approx - \frac{s}{H_0^2 \Omega_m}
        a \vec{\nabla} \phi_{\rm iso}.  
      \end{align}
      Using Eq.~\eqref{eq:phi_iso}, we see that
      $- \vec{\nabla} \cdot \vec{\psi} = \frac32 \gamma f_\bh s
      \delta_\bh^0$,
      which is identical to the isocurvature term in
      Eq.~\eqref{eq:delta_dm(a)}.
  
      To conclude, we have shown that the isocurvature contribution to
      the PDM density perturbations -- hence displacement field --
      given by the second term in Eq.~\eqref{eq:delta_dm(a)}, holds on
      both large scales and small scales. We only rigorously derived
      the adiabatic piece on scales larger than the inter-PBH
      separation, but argued that these scales dominate the adiabatic
      displacement field anyway. Moreover, the isocurvature mode soon
      dominates displacements, and all the more so on small scales.
    
    \end{subsubsection}

    \begin{subsubsection}{Implementation}
    
      We first need to choose the starting redshift of our
      simulations, $z_I \gg z_{\rm eq}$. It should be as high as
      possible without encountering any horizon scale effects.  We
      have selected $z_I=10^{6}$ as a good choice as the comoving
      horizon is $(aH)^{-1} \simeq 300$ kpc/$h$ $\gg L$ and there will
      not have been significant \dm{} accretion onto \bh{}. Indeed,
      the PDM bound mass is $\sim (1 - f_\bh) (a/a_{\rm eq})M_\bh$
      which, given our parameters, corresponds to around a single
      \dm{} particle:
      \begin{align}
        N_{\rm PDM, bound} \sim \frac{(1 - f_\bh)
        M_\bh s}{m_\dm} \sim \frac{10^6}{z}.  
      \end{align}

      \dm{} particles are initially created on a body-centered cubic
      lattice.  This lattice type, which we find necessary to obtain
      correct linear evolution, prevents the numerical growing mode
      from growing faster than the fluid limit
      \citep{bib:Joyce2005,bib:Marcos2008}.  \bh{} are not put on a
      lattice, but rather generated randomly throughout the simulation
      volume since they are expected to be Poisson distributed on
      these scales \citep{bib:AliHaimoud2018, Ballesteros_18,
        Desjacques_18}. Particles are then given a displacement
      $\vec{\psi}$ computed from the Zeldovich approximation
      \citep{bib:Zeldovich1970}
      $\vec{\psi} = - \vec{\nabla} \nabla^{-2} \delta_\dm$, where
      $\delta_\dm$ is given by Eq.~\eqref{eq:delta_pbh(a)} (note that
      we do not need to include $\delta_\bh^0$ for \bh{} as it is
      already generated by the initial random positions).  To be
      consistent with the Born approximation, we truncate the
      isocurvature displacement on scales of order the grid spacing.
      Since the \bh{} are not on the grid, we use a second order
      correction to the finite difference to prevent a self-force.
      Along a given dimension with cell index $i$, this is simply a
      Taylor expansion:
      \begin{align}
        \vec{\nabla}\phi &\rightarrow \frac{1}{2}(\phi_{i+1}-\phi_{i-1}) + {\rm dx}\left(\phi_{i+1}-2\phi_i+\phi_{i-1}\right)
      \end{align}
      where $|\rm{dx}|\le 1/2$ is the distance of the \bh{} from the
      center of cell $i$.
    
    \end{subsubsection}

  \end{subsection}

  \begin{subsection}{Gravitational Evolution and Accuracy}

    The main code reads in both sets of particles generated by the
    initial conditions.  Each particle is assigned a mass,
    \begin{align}
      m_i=\frac{\Omega_i}{\Omega_m}\frac{n_c^3}{N_i}
      \label{eq:simmass}
    \end{align}
    where $n_c=512$ is the number of grid cells, so that they
    contribute the correct energy density when interpolated to the
    grid. The two species of particles are distinguished via 1-byte
    particle identification numbers.  The background evolution is
    computed by solving Eq.~\eqref{eq:hubble} for the scale factor.

    \cpm{} has several accuracy parameters that can be changed from
    recommended values \citep{bib:HarnoisDeraps2013}.  The most
    important parameter for linear evolution is the range of the
    particle-particle (PP) force. We find that significant artificial
    growth on small scales occurs if the PP force is only computed
    over a single grid cell. This is due to deviations from the ideal
    $1/r^2$ force between particles due to the type of interpolation
    used.  Extending the PP force an additional $2$ cells results in
    much improved force accuracy (compare Figs.~7 and 14 of
    Ref.~\citep{bib:HarnoisDeraps2013}) and suppresses this numerical
    artifact. To demonstrate that we obtain accurate linear evolution,
    we show the power spectrum for the $f_\bh=0$ simulation in
    Fig.~\ref{fig:power1} which agrees acceptably well with linear
    evolution computed with \class~\cite{bib:Blas2011}. We also show
    the baryonic power spectrum at $a=10^{-2}$ in green, which is
    negligible relative to DM perturbations for most scales of
    interest.  Of course, nonlinear hydrodynamics could still be
    important.
    
    \begin{figure}
      \includegraphics[width=0.45\textwidth]{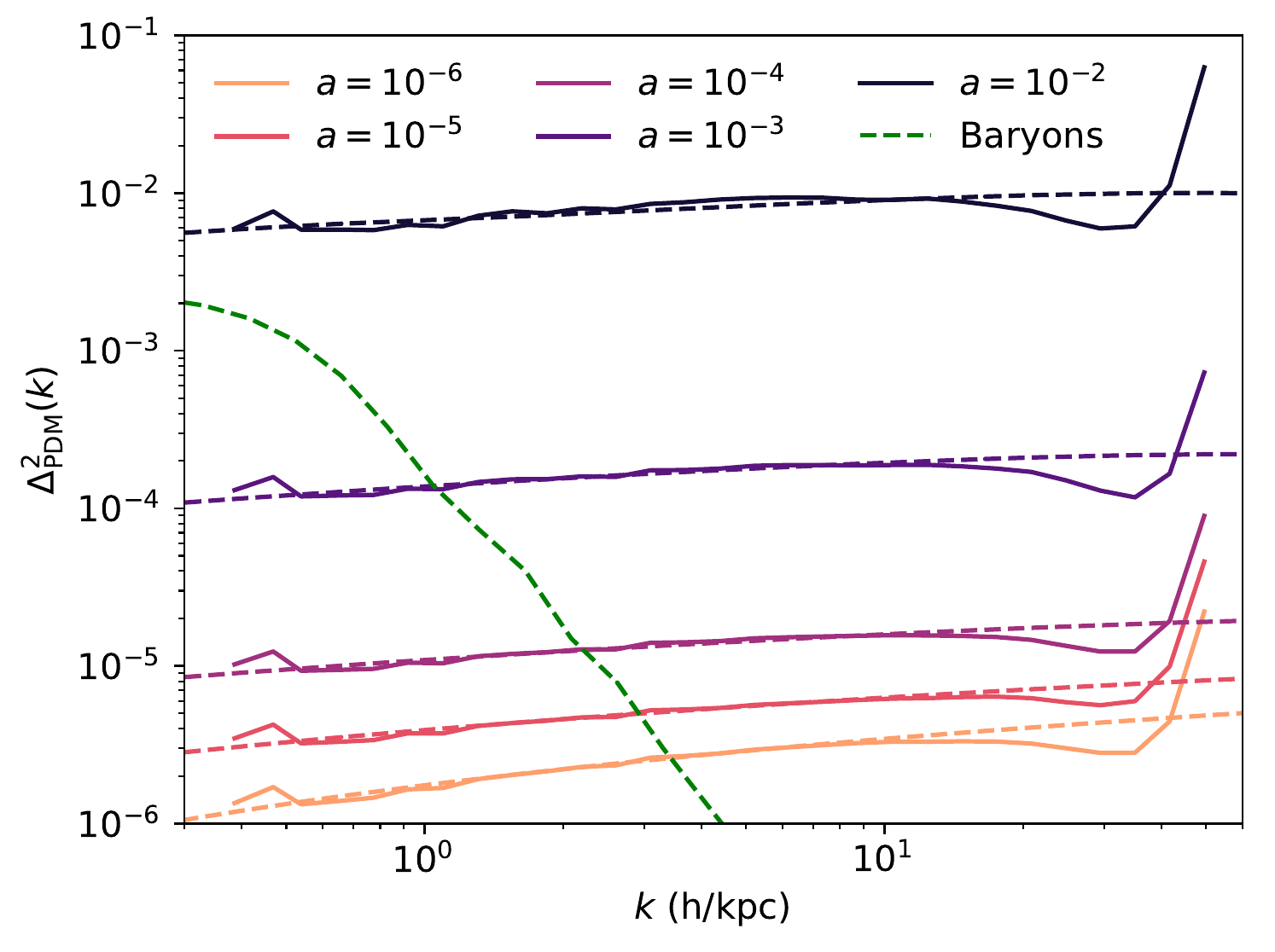}
      \caption{The PDM power spectrum as a function of scale factor,
        for $f_\bh = 0$.  Solid curves show the results from our
        simulation whereas dashed curves correspond to linear-theory
        results from \class~ \cite{bib:Blas2011}. Our simulations
        reproduce very well the growth of linear matter
        perturbations.  The green dashed
        curve is the baryonic power spectrum at $a=10^{-2}$, and
        illustrates that baryons remain mostly unclustered on all
        scales of interest, at least in the linear regime.}
      \label{fig:power1}
    \end{figure}
    
    We have also investigated the effects of the logarithmic time step
    limiter $ra_{\rm max}\ge da/(a+da)$, which prevents large jumps in
    redshift at early times, and the uniform offset applied each time
    step to avoid force artifacts when particles are outside the
    pp-force range due to the cubical rather than spherical particle
    search. We have increased the accuracy of both setting
    $ra_{\rm max}=0.005$ and allowing the offset to jump by up to 16
    fine cells.

    Lastly, we have tested the effects of gravitational softening in
    simulations containing a single \bh.  The default softening is
    that of a hollow sphere of radius $1/10$ of a fine grid cell.
    Ideally, for the \bh{} there would be {\it no} softening as they
    truly are discrete objects.  At great computational expense
    million particle $N$-body simulations with accurate binary orbital
    evolutions have been performed for globular clusters
    \citep{bib:Wang2015,bib:Wang2016}.  However, it is unclear whether
    this can be done in a cosmological context as binary orbits remain
    fixed whereas the Universe expands.  We therefore opt to soften
    the forces in our simulation.  Since strong force effects tend to
    evaporate halos by ejecting \bh, we expect our results to be an
    upper bound on \bh-\dm{} clustering although we caution that the
    softening length can have different effects in simulations with
    multiple species compared to traditional \dm{}-only ones. For
    instance, artificial scattering of \dm{} particles can be enhanced
    due to the large \bh{} mass creating an unwanted numerical heating
    effect \citep{bib:Dehnen2001} or artificial collisionality can
    occur and inhibit even linear evolution \citep{bib:Angulo2013}.
    We have explored the effects of various softening lengths in
    Appendix \ref{app:convergence} and have found the default
    softening length to provide acceptable results in acceptable time.

    We note a few potential issues with our evolution. The first
    one is non-spherical halo formation which occurs even if we turn
    off the adiabatic and isocurvature perturbations, i.e.~even if we
    start PDM particles exactly on the grid. This can be seen in the
    top left panel of Fig.~\ref{fig:slices} and in the zoom-in showing
    the \dm{} particles in Fig.~\ref{fig:zoom}. This is likely due to
    the radial orbit instability which afflicts $N$-body systems even
    with very carefully set spherically symmetric initial conditions
    (e.g.~\citep{bib:Vogelsberger2011}, who also show that power law
    profiles still form despite the asphericity). We do not expect
    this effect to significantly affect any of our results. A second
    issue is the fact that box-sized modes begin to go nonlinear at
    $z=99$ for $f_\bh\gtrsim 10^{-1}$ which could cause issues with
    power transfer from large to small scales \citep{bib:Bagla1997}.
    Lastly, for simulations with $f_\bh \ll 1$, the assumed
    periodicity of the simulation becomes artificial as we are not
    well sampling Poisson fluctuations on scales larger than the box
    size. This will introduce sample variance into our results that
    could affect quantities such as the halo mass function and perhaps
    \bh{} velocities when $f_\bh\ll1$, but should not affect halo
    profiles.
   
    \begin{figure}
      \includegraphics[width=0.45\textwidth]{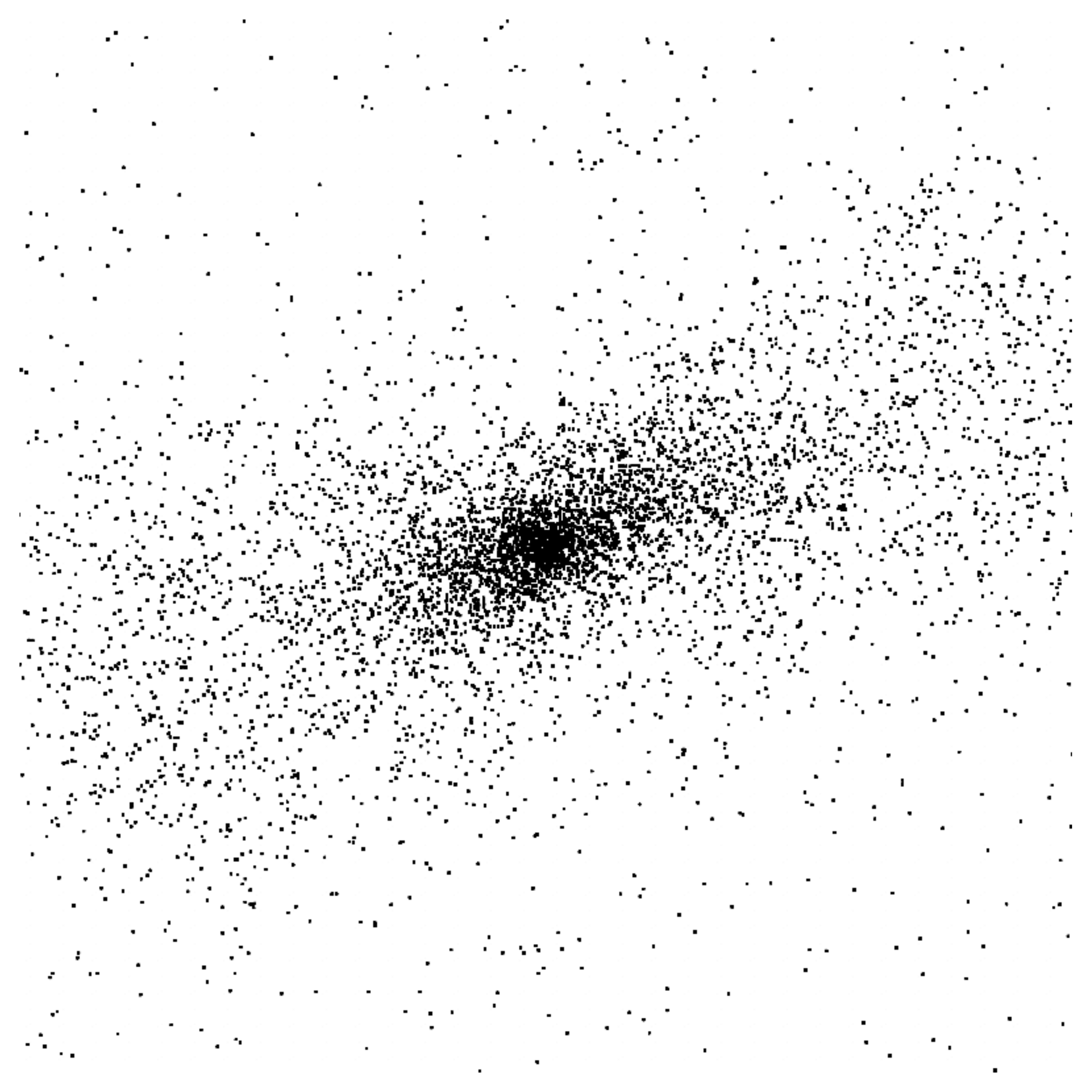}
      \caption{Zoom-in showing the \dm{} particles around the single
        PBH in our $f_\bh=10^{-5}$ simulation (including adiabatic
        perturbations). Particles shown are in $(1/64)^3$ of the
        simulation box around the \bh. The PDM distribution is clearly
        aspherical in the outskirts, likely due to the radial orbit
        instability.}
      \label{fig:zoom}
    \end{figure}
   
  \end{subsection}

  \begin{subsection}{Halofinder}

    To understand gravitational clustering, we require estimates of
    how much \dm{} and \bh{} are bound in nonlinear halos. \cpm{} has
    a run-time spherical overdensity halo finder. It first finds peaks
    in the \dm{} density field and then searches radially until the
    mean overdensity within $r$, $\Delta=\rho_\dm(<r)/\bar{\rho}_\dm$,
    drops to $\Delta_{\rm vir}\simeq18\pi^2$. This defines the virial
    radius $r_\hl$. Note that the value of $\Delta_{\rm vir}$ has not
    been corrected for the presence of radiation, which is percent
    level at $z=99$.  It also does {\it not} consider the \bh{}
    particles since only the \dm{} particles will have a smooth
    density field.  Particles are not allowed to be in more than one
    halo, and halos are required to have at least $100$ \dm{}
    particles to be included in the catalogue.  The halofinder is
    parallelized with the same volume decomposition as the main code,
    and therefore processes can find halos located outside their local
    volume.  We do not include any such halos in order to avoid
    duplicates.  We use the resulting halo virial masses as our
    estimate of the \dm{} accretion.  Lastly, we note that at higher
    $f_\bh$ the halofinder begins to find a small fraction of halos
    without \bh{} and that there are also many \bh{} that are not
    considered part of halos.  This is certainly artificial, since
    even a perfectly isolated \bh{} should be called a halo.  In the
    future, designing a dedicated \lpbh{} halofinder could be
    warranted.
   
  \end{subsection}

\end{section}

\begin{section}{Results}

  \begin{subsection}{Matter power spectrum}
    \label{sec:P(k)}

    \begin{figure}
      \includegraphics[width=0.45\textwidth]{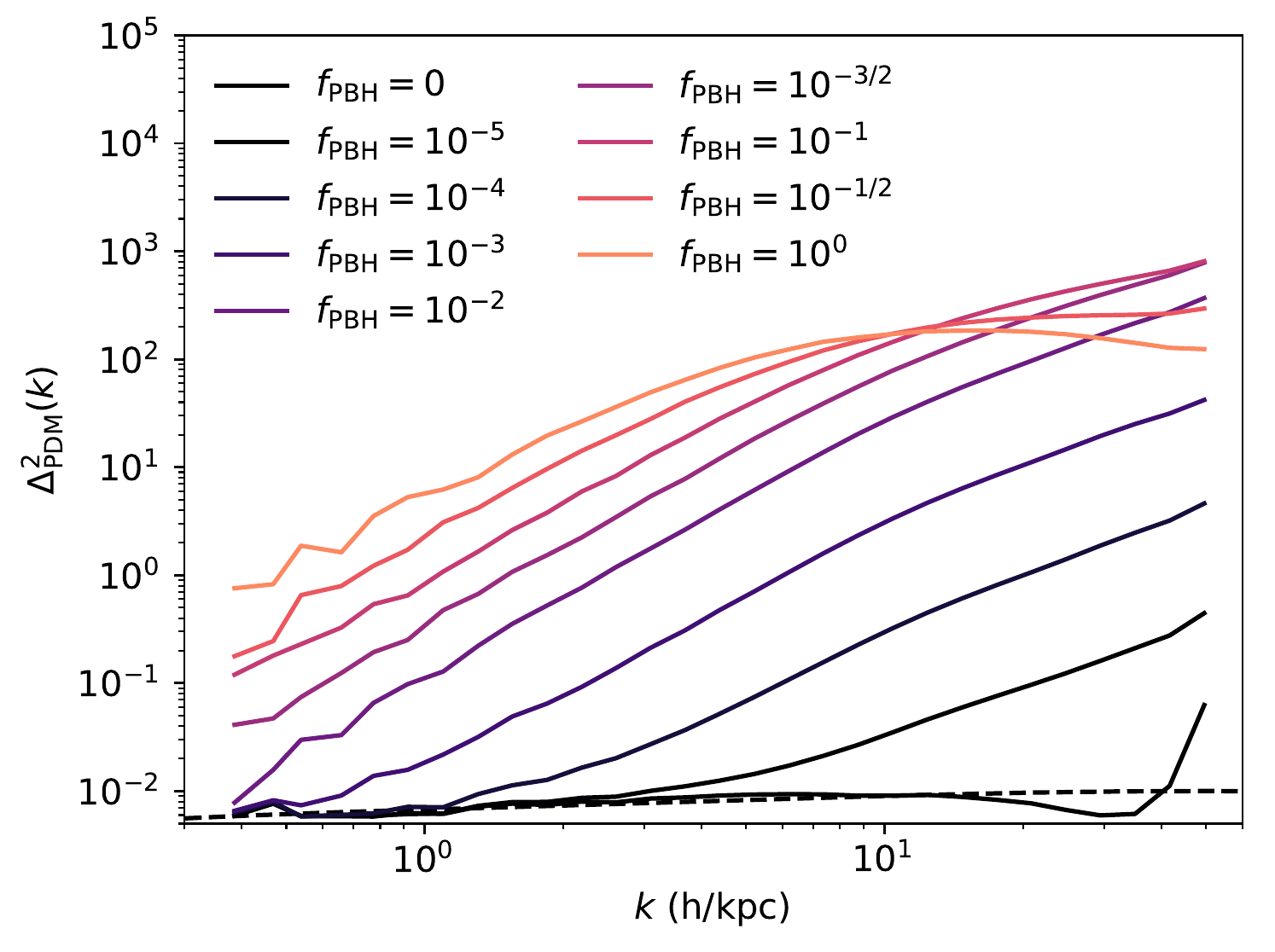}
      \caption{The \dm{} power spectrum as a function of $f_\bh$ at
        $a=10^{-2}$.  The presence of \bh{} causes significant
        nonlinear growth not found in standard \lcdm.}
      \label{fig:power}
    \end{figure}

    \begin{figure}
      \includegraphics[width=0.45\textwidth]{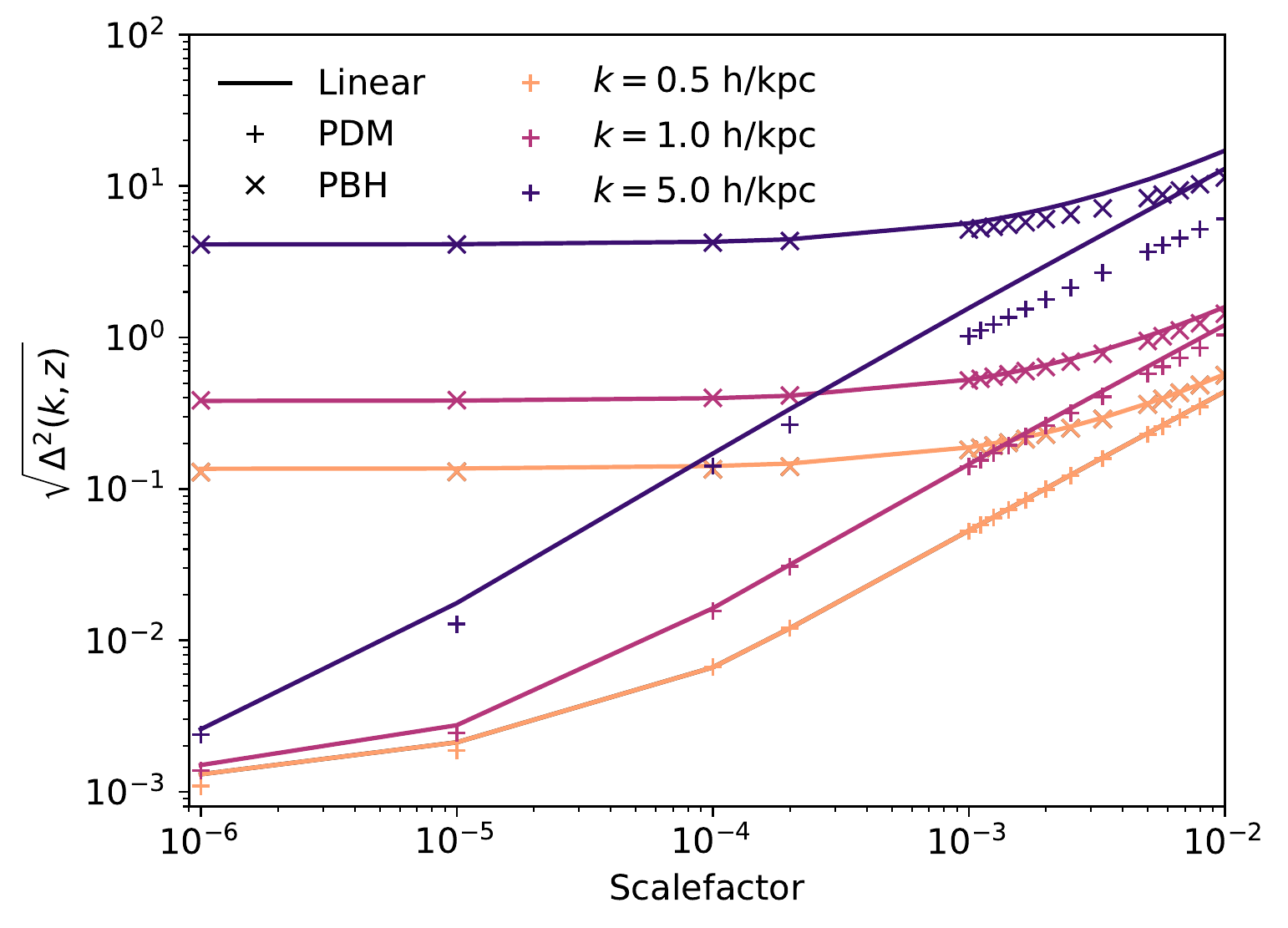}
      \caption{Growth of the PDM and PBH perturbation as a function of
        scale factor, for $f_{\bh} = 0.1$, and for several wavenumbers
        $k$. Linear theory matches the numerical result well, as long
        as $\Delta^2_\bh \lesssim 1$.}
      \label{fig:modes}
    \end{figure}

    We start by computing the \dm{} power spectrum which we show in
    Fig.~\ref{fig:power}.  The black lines (solid is simulation,
    dashed is \class{}) show linear evolution of the \dm{} when there
    are no \bh{} ($f_\bh=0$).  The other curves show the resulting
    power spectra at $z=99$ as a function of $f_\bh$.  Increasing
    $f_\bh$ disrupts the linear evolution at larger and larger scales,
    and completely dominates on all scales of the box for
    $f_\bh \gtrsim 10^{-2}$.  Interestingly, the \dm{} power spectrum
    turns over at large $k$ for high $f_\bh$. This is likely due to
    the fact that halo profiles are very cuspy around isolated PBHs
    but significantly less so for halos containing multiple PBHs. The
    latter are more common as $f_\bh$ is increased.

    We next investigate the evolution of individual modes. We show the
    growth of density perturbations for several wavenumbers from the
    $f_\bh=0.1$ simulation in Fig.~\ref{fig:modes} for both \dm{} and
    \bh.  We compare this result to the linear evolution of PBH and
    PDM perturbations, derived in Section \ref{sec:pert-theory}. Note
    that linear theory only applies on scales where both $\delta_\dm$
    and $\delta_\bh$ are small. In particular, it only applies for
    scales larger than the mean PBH separation. Using
    Eqs.~\eqref{eq:delta_PDM-lin}- \eqref{eq:delta_PBH-lin} with
    $T_{\rm iso}(a) = D_+(a/a_{\rm eq})$ given by
    Eq.~\eqref{eq:D+-approx}, and the fact that primordial adiabatic
    and isocurvature perturbations are uncorrelated, we find the
    following power spectra:
    \begin{align}
      \Delta^2_{\dm}(a) &= \Delta_{\rm ad}^2(a) + f_\bh^2 \left(D_+(a) - 1 \right)^2 (\Delta_{\bh}^0)^2~~~\\
      \Delta^2_\bh(a) &= \Delta_{\rm ad}^2(a) \nonumber\\
                        &+ \left[ 1+ f_\bh (D_+(a) - 1)\right]^2 (\Delta_\bh^0)^2,
    \end{align}
    where $\Delta_{\rm ad}^2(a)$ is the power spectrum of the pure
    adiabatic mode at scale factor $a$, $(\Delta_\bh^0)^2$ is the
    primordial PBH (Poisson) power spectrum and we approximated
    $\delta_{\rm iso}^0 \approx \delta_\bh^0$.
      
    We compare these results with those of our simulations in
    Fig.~\ref{fig:modes}. We see that they agree quite well for linear
    modes, but increasingly differ as nonlinear evolution occurs, as
    can be expected. Interestingly, non-linear growth is \emph{slower}
    than linear theory predicts.

  \end{subsection}

  \begin{subsection}{Halo properties}
    We begin by quantifying how much \dm{} and \bh{} are found within
    halos.  We expect this to depend sensitively on $f_\bh$.  For
    small $f_\bh$, PBHs will mostly be isolated from one another, and
    so halos can only increase mass by accumulating more and more \dm.
    On the other hand, for higher $f_\bh$, the \bh{} themselves can
    become bound to one another, yielding potentially much larger
    halos with very different internal structures.

    \begin{subsubsection}{\dm{} clustering in halos}

      \begin{figure}
        \includegraphics[width=0.45\textwidth]{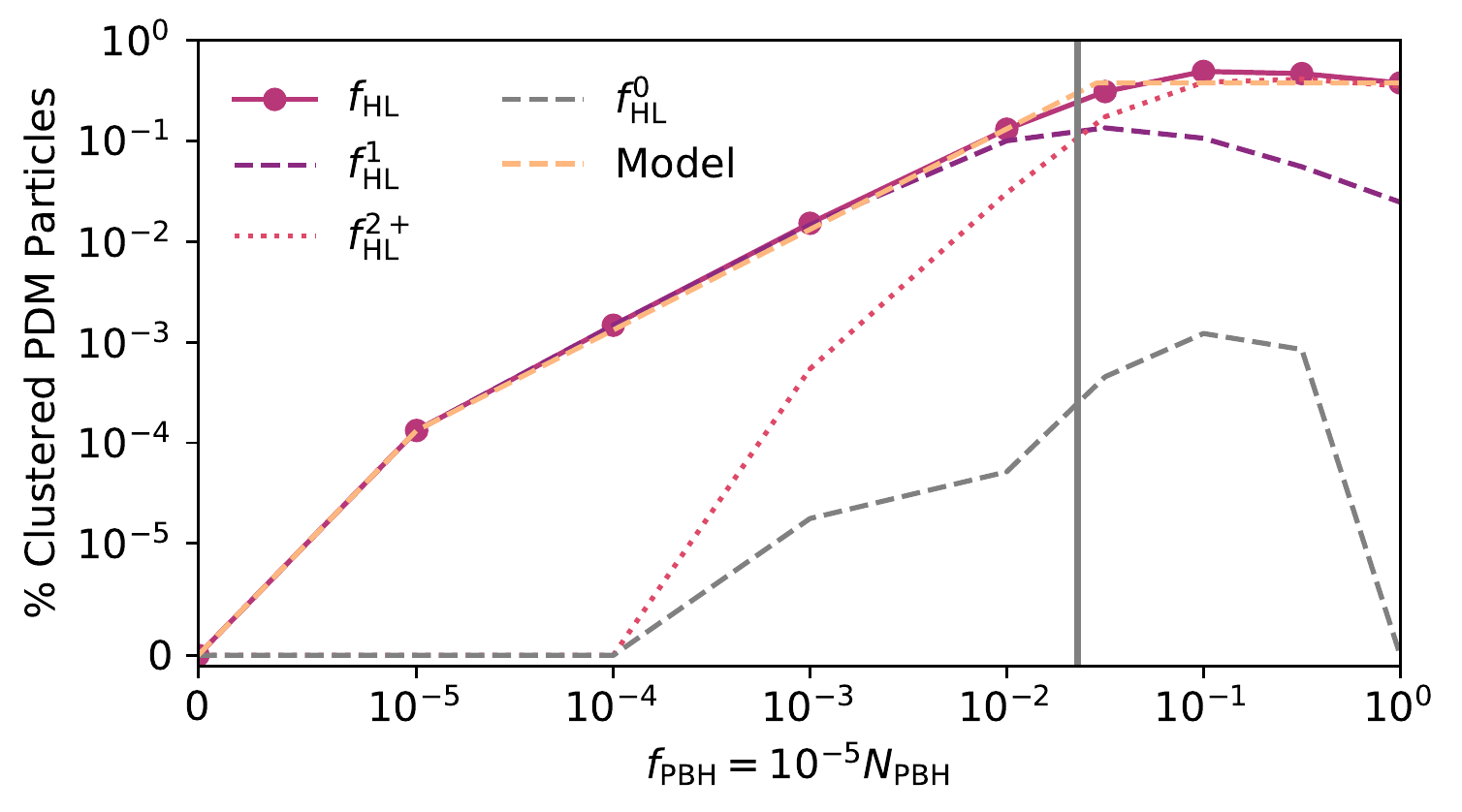}
        \includegraphics[width=0.45\textwidth]{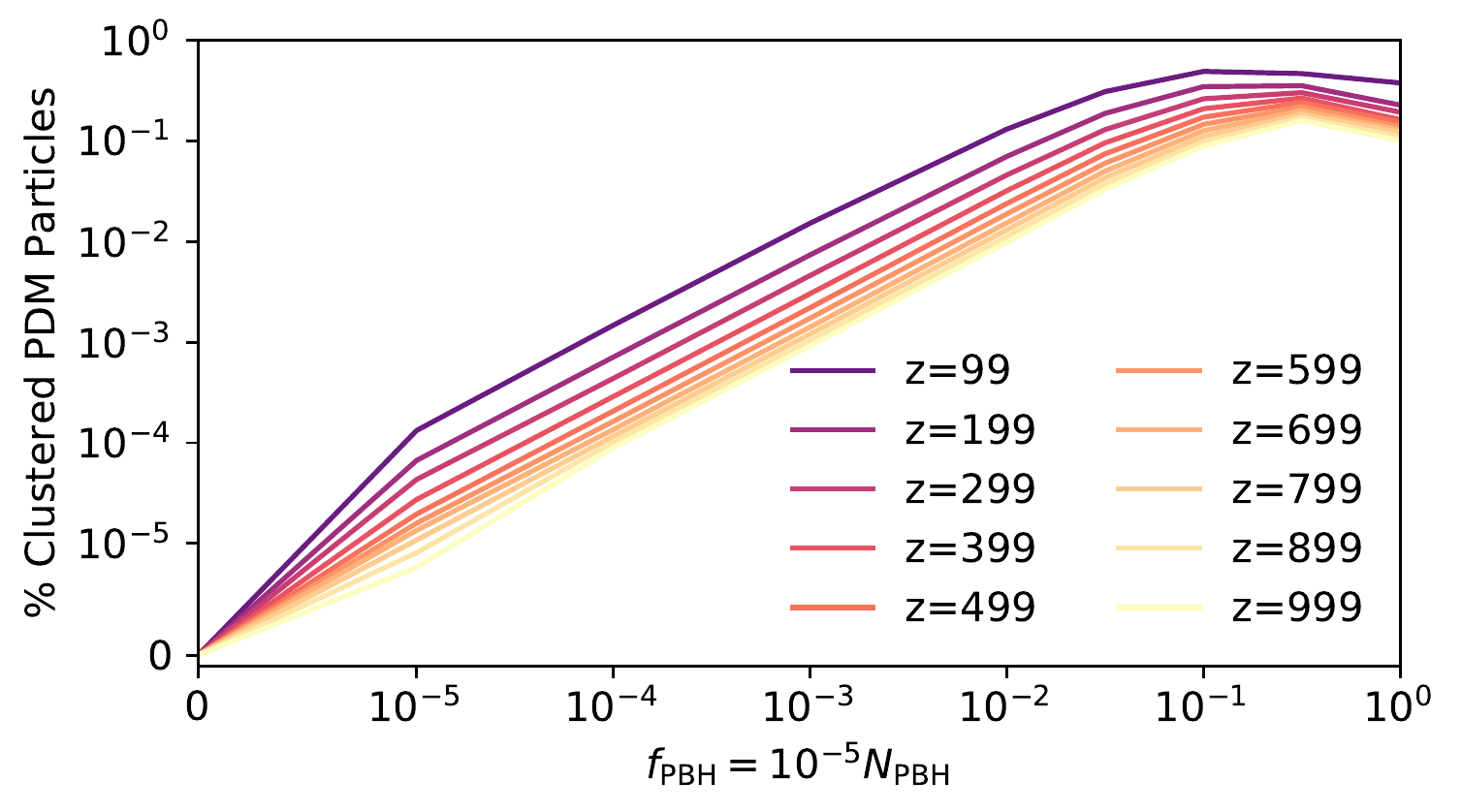}
        \caption{{\it Top.} Fraction of PDM contained within halos at
          $z = 99$. The solid purple line shows the fraction of PDM in
          all halos. The dashed and dotted curves indicate the amount
          which is in halos with only one \bh{} and those with
          multiple PBHs, respectively. The grey vertical line is a
          prediction for their intersection. The dashed grey curve is
          the fraction of halos found without \bh{}.  The orange curve
          shows a simple interpolating model. {\it Bottom.}  The same
          fraction shown at a variety of redshifts.}
        \label{fig:hl_mass_fr}
      \end{figure}

      We begin by computing the fraction of \dm{} particles that are
      clustered as a function of $f_{\bh}$ which is given by
      $f_\hl=N_{\dm \in \hl}/N_\dm$ where $N_{\dm \in \hl}$ is the
      number of \dm{} particles in halos.  We show the results at
      $a=10^{-2}$ in the top panel of Fig. \ref{fig:hl_mass_fr},
      alongside the fraction of \dm{} in halos without \bh{}
      ($f_\hl^0$).  At $f_\bh=0$ we find no halo as expected.  We
      furthermore break $f_\hl$ based on whether it comes from a halo
      with one ($f_\hl^1$) or multiple ($f_\hl^{2+}$) \bh.  There is a
      clear transition between isolated \bh{} ($f_\hl^{1}$) and
      clustered \bh{} ($f_\hl^{2+}$) that occurs between
      $10^{-2}<f_\bh<10^{-1}$.  To quantify this transition, we
      require knowledge of how \bh{} are distributed amongst halos of
      varying sizes, i.e.~the halo mass function.  A theoretical
      prediction, calculated in the next section, is shown as the grey
      vertical line, and agrees reasonably well with the numerical
      results.

      This figure can be understood straightforwardly by considering
      limiting cases. For $f_\bh \ll 1$, we expect halos to be well
      separated on average and accrete identically, so
      $f_\hl = c_0 f_\bh$.  For $f_\bh\sim1$, we expect the majority
      of the \dm{} to be nonlinear, however only some fraction will be
      within halo virial radii and so $f_\hl = c_1$ when
      $\alpha \gg 1$.  We show a simple interpolation between these
      limits:
      \begin{align}
        f_\hl \simeq {\rm min}\left[ c_0 f_\bh,c_1\right]
      \end{align}
      where $c_0$ is the ratio of \dm{} halo fraction to \bh{}
      fraction for $f_\bh=10^{-5}$ and $c_1$ is the ratio for
      $f_\bh=10^0$.  In the bottom panel of Fig.~\ref{fig:hl_mass_fr}
      we show the evolution of $f_\hl$ as a function of redshift. We
      tabulate the values of $c_0$ and $c_1$ in
      Table.~\ref{tab:fhl}. The coefficient $c_0$ scales approximately
      linearly with scale factor, as expected from the fact that the
      mass of PDM bound to a PBH is proportional to
      $(1 - f_\bh) M_\bh (a/a_{\rm eq})$ (see \cite{bib:Mack2007} and
      Section \ref{sec:small-scales}).
        
      \begin{center}
        \begin{table}
          \begin{tabular}{r|c|c}
            $z$ & $c_0$ & $c_1$\\
            \hline
            $999$ & 0.79 & 0.10\\
            $299$ & 4.3 & 0.19\\
            $99$ & 13 & 0.38\\
          \end{tabular}
          \caption{Coefficients to compute the fraction of \dm{}
            particles in halos: $f_\hl={\rm min}[c_0 f_\bh, c_1]$ as a function of redshift.}
          \label{tab:fhl} 
        \end{table}
      \end{center}
        
      Understanding $f_\hl$ gives a straightforward way to estimate
      the halo mass as a function of the number of \bh{} it contains.
      Let $N_{\bh/\hl}$ and $N_{\dm/\hl}$ be the number of \bh{} and
      \dm{} particles in a halo.  By definition, its mass is given by:
      \begin{align}
        M_\hl&= N_{\bh/\hl}m_\bh + N_{\dm/\hl}m_\dm \\
             &= N_{\bh/\hl}m_\bh\left[1+\frac{N_{\dm/\hl}}{N_{\bh/\hl}}\frac{m_\dm}{m_\bh}\right].
      \end{align}
      If we now assume that $N_{\dm/\hl}$ is directly proportional to
      $N_{\bh/\hl}$ and that all \bh{} are in halos, we can obtain
      $N_{\dm/\hl}= N_{\bh/\hl}
      \frac{N_\dm}{N_\bh}\frac{f_\hl}{1-f_\bh}$
      (this is equivalent to assuming
      $N_{\dm/\hl}=N_{\dm\in\hl}/N_\hl$ and
      $N_{\bh/\hl}=N_\bh/N_\hl$). Using this relation the halo mass
      becomes:
      \begin{align}
        \label{eq:halomass}
        M_\hl  &\simeq N_{\bh/\hl}m_\bh\left[1+f_\hl\frac{1-f_\bh}{f_\bh}\right]. 
      \end{align}
      We show this result in Fig.~\ref{fig:hl_mass} and find the
      approximation is accurate to a factor of a few.
    
      \begin{figure}
        \includegraphics[width=0.45\textwidth]{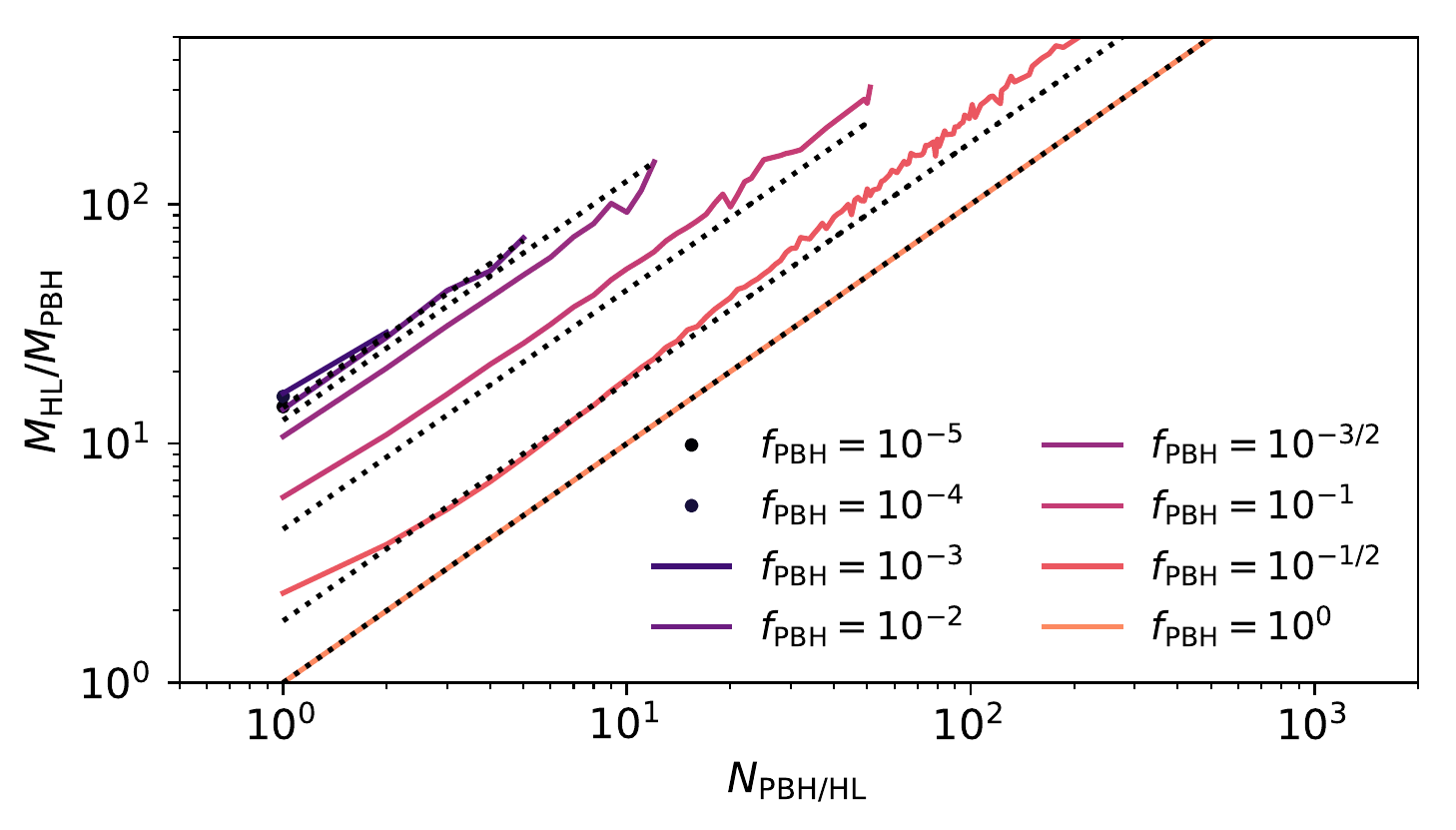}
        \caption{Halo mass as a function of the number of \bh{}
          contained within the halo virial radius.  A simple
          interpolating model is also shown as dotted lines.}
        \label{fig:hl_mass}
      \end{figure}

    \end{subsubsection}

    \begin{subsubsection}{Halo Mass Function}

      \begin{figure}
        \includegraphics[width=0.45\textwidth]{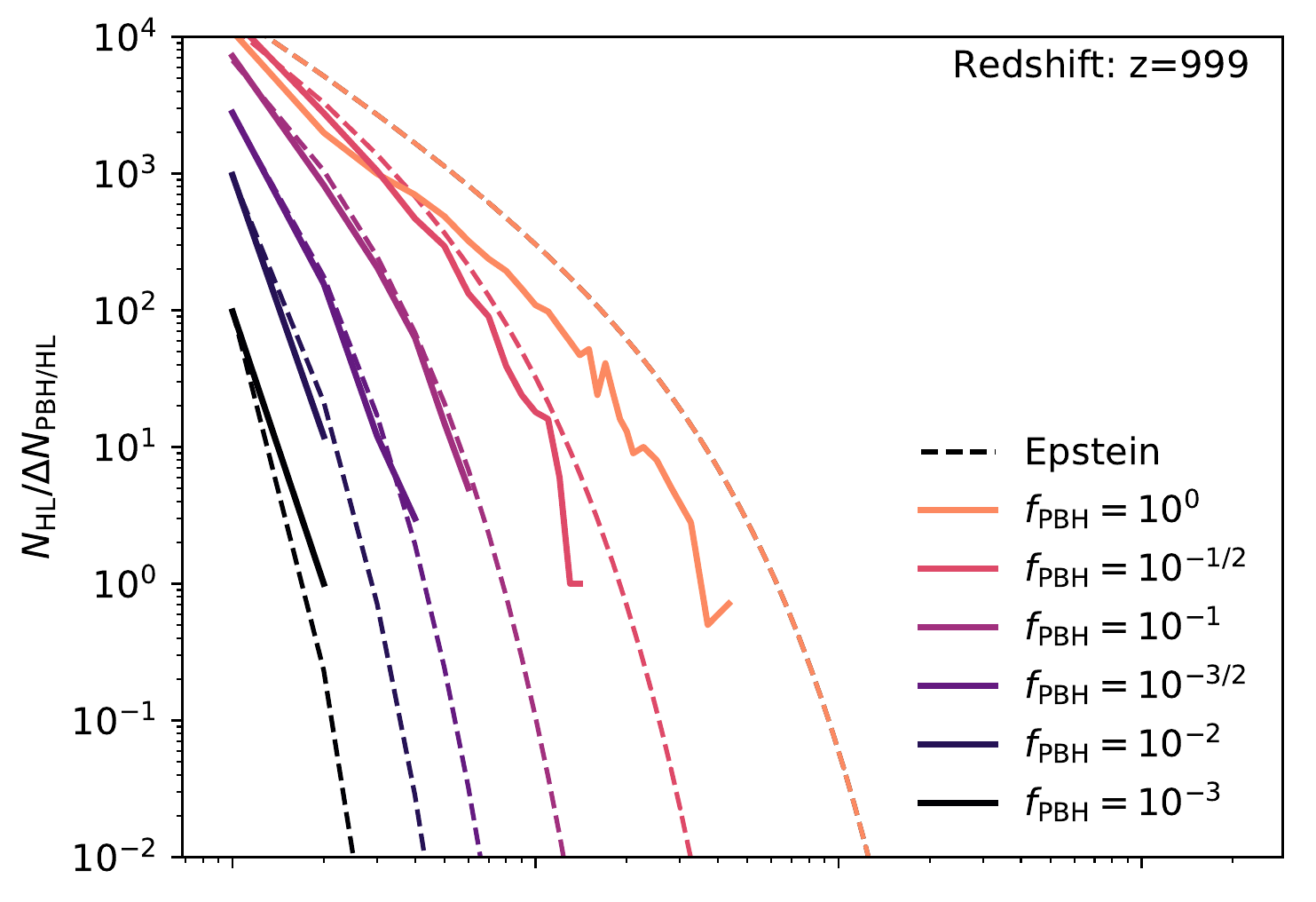}\\
        \includegraphics[width=0.45\textwidth]{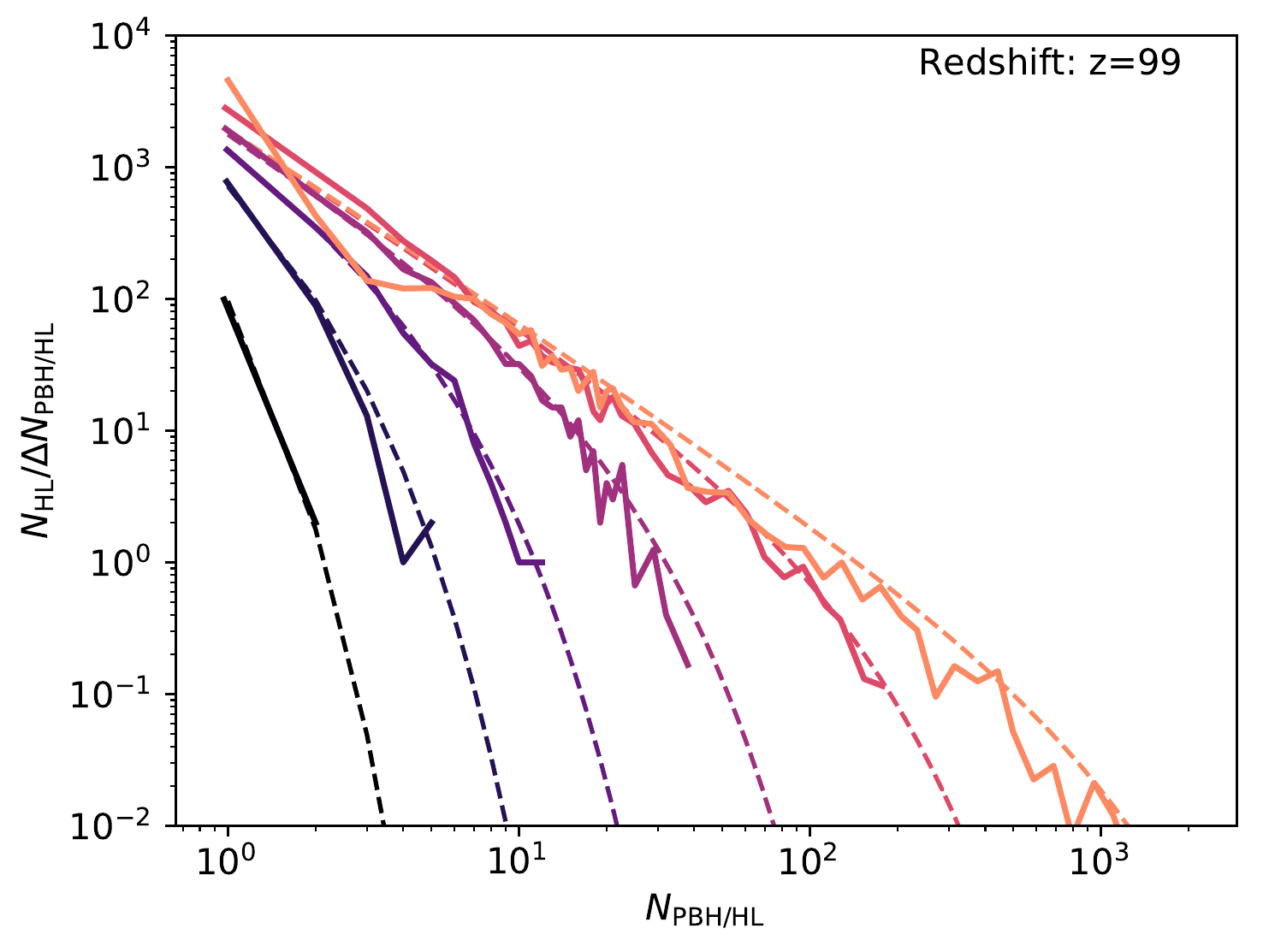}
        \caption{The abundance of halos containing a given number of
          \bh{}, which we call the halo mass function. Solid lines are
          determined from $N$-body simulations.  Dashed lines are
          theoretical predictions assuming Poisson statistics.}
        \label{fig:hl_number_fn}
      \end{figure}

      In order to determine when halo growth occurs via
      quasi-spherical accretion or mergers, we need to be able to
      determine how many \bh{} are isolated as a function of $f_\bh$.
      To do this, we compute the number of halos containing a given
      number of \bh, $N_\hl(N)$, where, for notational simplicity, we
      use $N=N_{\bh/\hl}$ in this discussion.  As we have shown in
      Fig.~\ref{fig:hl_mass}, the halo mass is nearly linearly related
      to the number of \bh{} it contains.  We will therefore refer to
      $N_\hl(N)$ as the halo mass function.  For initially
      Poisson-distributed particles, \citet{bib:Epstein1983} computed
      the exact distribution arising from an initial density
      $\delta_\bh^I> \delta_*$ as:
      \begin{align}
        \label{eq:Epstein}
        N_\hl(N) &=\frac{N_\bh}{N} \frac{\delta_*}{1+\delta_*}
                   \left(\frac{N}{1+\delta_*}\right)^{N-1}
                   \frac{\exp\left[-\frac{N}{1+\delta_*}\right]}{ (N-1)! } \nonumber\\
                 &= \delta_* \frac{N_{\bh}}{N} \frac{(N/e)^N}{N!} \textrm{e}^{- N/N_*}\nonumber\\
        N_* &\equiv \left(\log(1+\delta_*) - \frac{\delta_*}{1+\delta_*}
              \right)^{-1}, 
      \end{align}
      where $N_\bh = 10^5 f_\bh$ is the total number of PBHs.
        
      For $N \gg 1$, we may use Stirling's approximation for the
      factorial and obtain
      \begin{align}
        N_\hl(N) \approx& \frac{\delta_* ~N_\bh}{\sqrt{2 \pi} N^{3/2}} \textrm{e}^{- N/N_*}. \label{eq:Epstein-2}
      \end{align}
      The fractional error of this approximation is $0.08/N$,
      \emph{independent} of $\delta_*$; this approximation is
      therefore accurate to better than 10\% even for $N \sim 1$.
        
      When $\delta_*\ll 1$, we may Taylor-expand
      $N_*^{-1} \approx \delta_*^2/2 + \mathcal{O}(\delta_*^3)$.
      Provided $N \ll \delta_*^{-3} \sim N_*^{3/2}$, we may neglect
      terms of order $N \delta_*^3$ in the exponent and recover
      \citep{bib:Epstein1983} the Press-Schechter function
      \citep{bib:Press1974}
      \begin{align}
        \label{eq:press_schechter}
        N_\hl(N) \approx \frac{\delta_* ~N_\bh}{\sqrt{2 \pi} N^{3/2}} 
        \exp\left[ -\frac{\delta_*^2}{2}N\right].
      \end{align}
      In practice, this approximation always breaks down for
      sufficiently large $N$ and we only use Eqs.~\eqref{eq:Epstein}
      or \eqref{eq:Epstein-2}.
      
      For a given scale factor $a$, the minimum initial PBH
      overdensity $\delta_*$ is determined as follows. We require that
      the initial total CDM overdensity
      $\delta_{c}^I = f_{\bh} \delta_*$ has collapsed into a halo by
      scale factor $a$ (this assumes negligible fluctuations in the
      PDM component). This is equivalent to requiring that the
      linearly-extrapolated CDM overdensity
      $\delta_{\rm lin}(a) = D_+(a) \delta_c^{I}$ has reached a
      critical value $\delta_{\rm cr}(a)$, where $D_+$ is computed in
      Section \ref{sec:P(k)}. This implies
      \begin{align} \delta_*(a) = \frac{\delta_{\rm cr}(a)}{D_{+}(a)
          f_{\bh}}. \label{eq:deltai-mat}
      \end{align}
      When collapse occurs well inside matter domination, and when
      baryons cluster like dark matter, the critical density is
      $\delta_{\rm cr} = 1.69$. However, it can differ significantly
      from this value as collapse occurs closer to matter-radiation
      equality, and on scales where baryons remain unclustered. We
      explicitly compute $\delta_{\rm cr}(a_{\rm coll})$ in Appendix
      \ref{app:spherical-collapse}. We find, for instance, that
      $\delta_{\rm cr} \approx 2.07$ at $z = 999$, and even at
      $z = 99$, $\delta_{\rm cr} \approx 1.71$. The minimum initial
      PBH overdensity is therefore $\delta_* \approx 0.43/f_\bh$ and
      $0.052/f_\bh$ at $z = 999$ and 99, respectively.  We find that
      the Epstein mass function \eqref{eq:Epstein}, with $\delta_i$
      given by Eq.~\eqref{eq:deltai-mat}, give a good match to our
      halo mass function at $z = 99$, see lower panel of
      Fig.~\ref{fig:hl_number_fn}.  The Epstein function also matches
      our halo mass function reasonably well at $z = 999$ for
      $f_\bh \lesssim 10^{-1/2}$, see upper panel of
      Fig.~\ref{fig:hl_number_fn}. The poorer match at $f_\bh = 1$
      could be due to our halo finder, which is exclusively based on
      PDM particles and misses some \bh.
        
      Given the number function, Eq.~\eqref{eq:Epstein}, we can now
      compute the value of $f_\bh$ for which halo formation
      transitions from the seed to the Poisson mechanism.
      Specifically, this is when half the \bh{} are in halos with 1
      PBH, or
      \begin{align}
        \frac{N_{\hl}(1)}{N_\bh} =
        \frac{\delta_*}{1+\delta_*}\exp\left[\frac{-1}{1+\delta_*}\right]
        =\frac{1}{2}.  
      \end{align}
      This is satisfied for $\delta_*=2.175$ so
      $f_\bh\simeq0.02(1+z)/100$.  This prediction is shown in
      Fig.~\ref{fig:hl_mass_fr} as the vertical grey line and matches
      the numerical result quite well.

    \end{subsubsection}

    \begin{subsubsection}{Halo Profiles}

      We now consider the \dm{} density profiles that form around the
      halos in the simulation.  We start by considering the single
      isolated \bh{} in the $f_\bh=10^{-5}$ simulation and show its
      profile as a function of redshift in Fig.~\ref{fig:profile_z}.
      At early times ($a=10^{-3}$), we find a profile with power law
      slope of $-2.28\pm0.08$ which is consistent with both the
      theoretical prediction ($-2.25$) and other numerical simulations
      \cite{bib:Adamek2019}.  At later times the profile becomes
      steeper, with a slope of $-2.55\pm0.04$ at $a=10^{-2}$.

      \begin{figure}
        \includegraphics[width=0.45\textwidth]{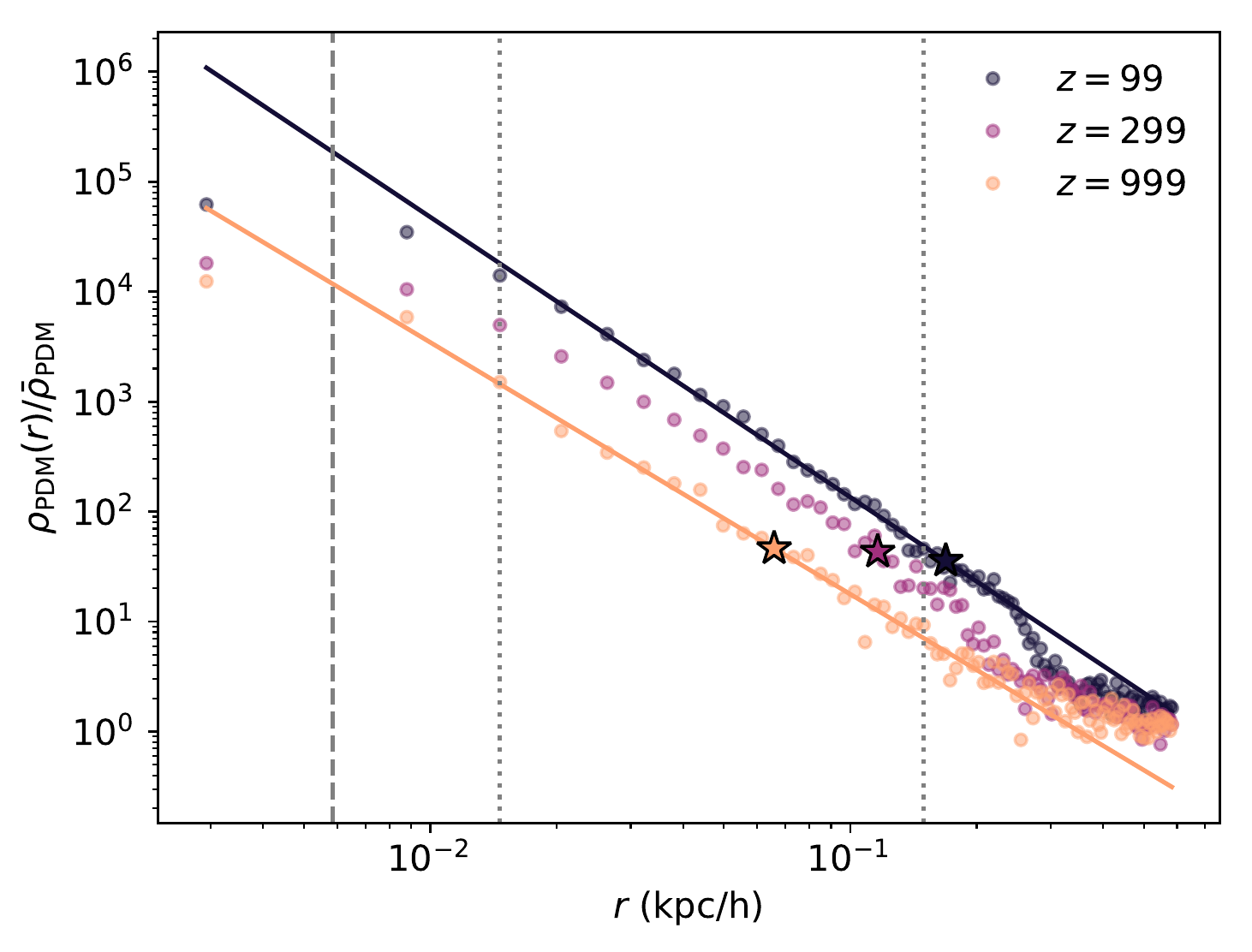}
        \caption{Spherically averaged \dm{} density profiles around the single
          black hole in the $f_\bh=10^{-5}$ simulation as a function
          of comoving radius, and for several redshift.  We find that
          the profile features a power law at all times, which
          steepens as time progresses. Best-fit power laws at
          $a=10^{-3}$ and $a=10^{-2}$ are shown as solid lines.  The
          dashed grey curve indicates the force softening length
          whereas the dotted grey curves indicate the radii included
          in the power law fit.  The stars indicate the halo virial
          radii.}
        \label{fig:profile_z}
      \end{figure}
      
      Next, we investigate how the \dm{} profiles vary with $f_\bh$ at
      $a=10^{-2}$.  We identify halos containing only a single \bh{}
      and compute the average \dm{} density profile of all such halos.
      We show the result in Fig.~\ref{fig:profile_f}, with error bands
      showing the standard deviation. We find that a power law profile
      is retained, regardless of $f_\bh$ but that the slope becomes
      steeper with increasing $f_\bh$, up to a value of $-3.24\pm0.04$
      for $f_\bh=1$. We note that we slightly reduced the fit range
      since the halo radii are significantly smaller at higher
      $f_\bh{}$.
 
      \begin{figure}
        \includegraphics[width=0.45\textwidth]{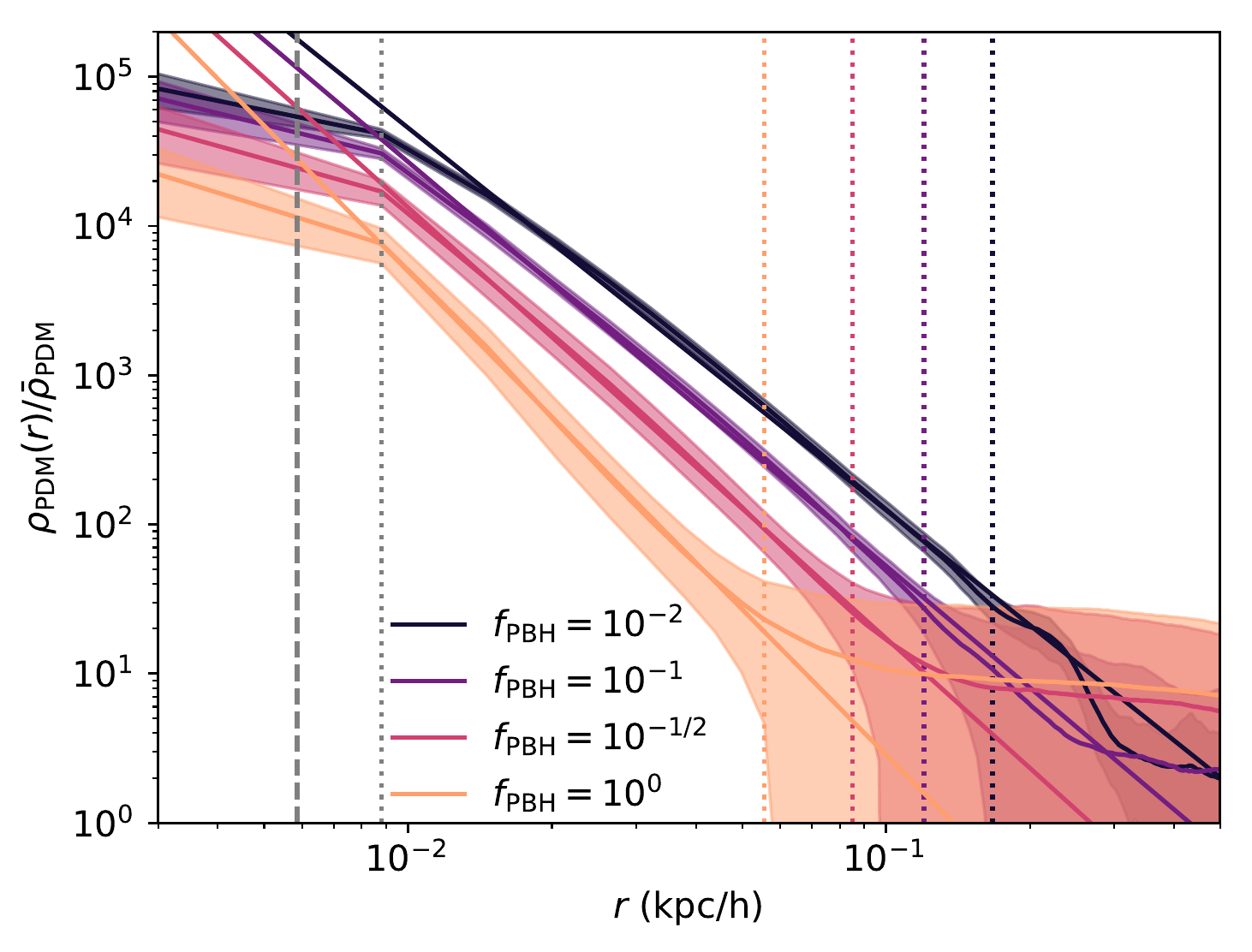}
        \caption{Average \dm{} density profiles around halos
          containing a single \bh{} at $a=10^{-2}$, as a function of
          $f_\bh$, with error bands showing the standard deviation. We
          find that PDM halos around isolated \bh{} have power law
          profiles, whose slope steepens with increasing $f_\bh$. The
          dashed line shows the softening length, whereas dotted lines
          indicate the fit region, with largest value being the halo
          virial radius.}
        \label{fig:profile_f}
      \end{figure}

      Finally, we consider halos containing more than one \bh{}.  As
      an example, we show in Fig.~\ref{fig:profile_13_17} the mean
      density profile of halos containing between 13 and 17 \bh.  We
      find the profile to be significantly shallower.  While it is
      suggestive that the profiles are more cored than power law, we
      caution that softening is likely playing a role and higher
      resolution simulations are necessary to determine if the profile
      turns over.
    
      \begin{figure}
        \includegraphics[width=0.45\textwidth]{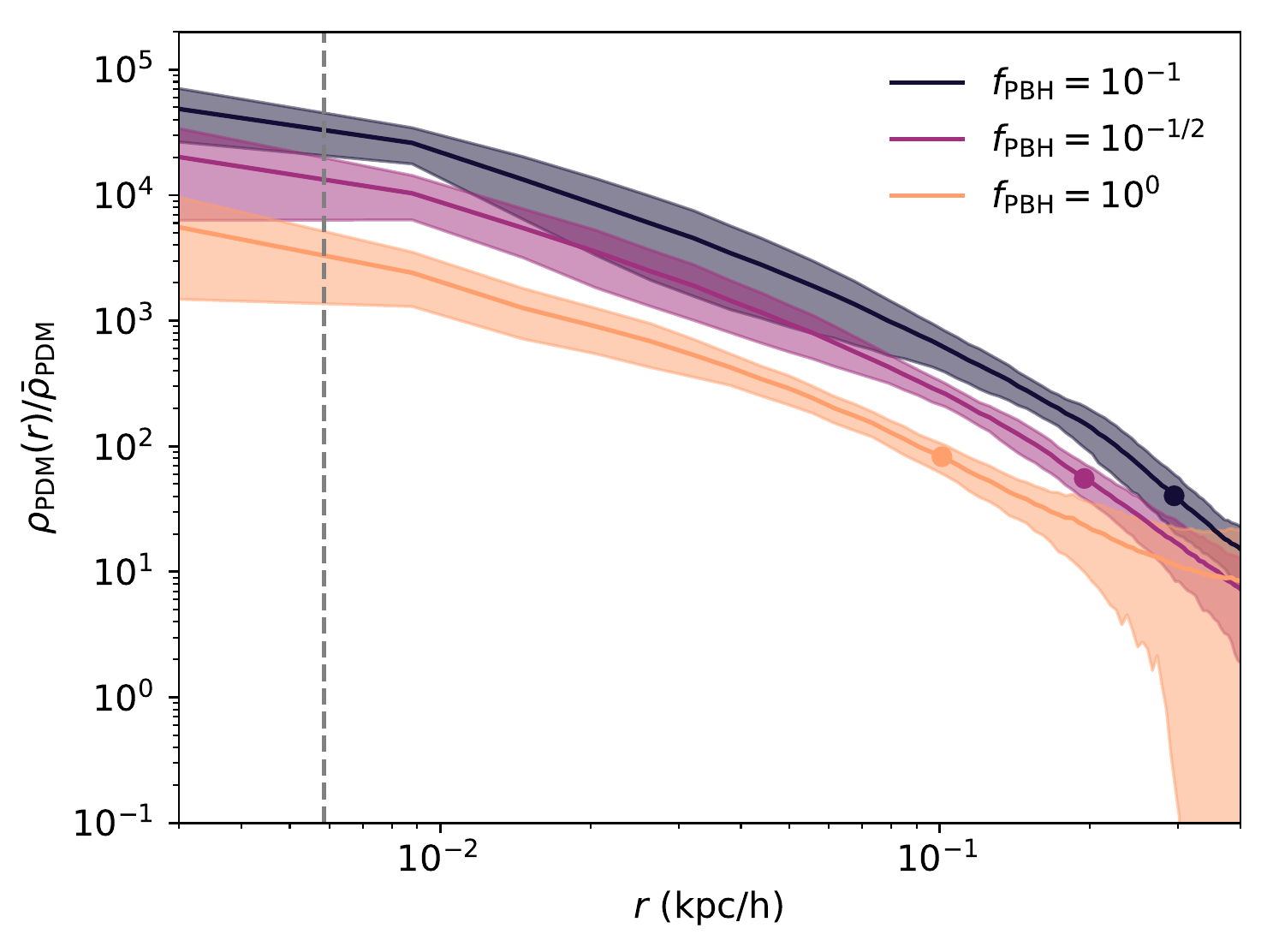}
        \caption{\dm{} density profiles around halos containing
          between 13 and 17 \bh{} at $a=10^{-2}$.  The profiles are
          significantly shallower than the power laws seen in
          Fig.~\ref{fig:profile_f}.}
        \label{fig:profile_13_17}
      \end{figure}

    \end{subsubsection}

  \end{subsection}

  \begin{subsection}{Nonlinear \bh{} Velocities}

    We now turn our attention to the PBH velocity
    distribution. Determining this quantity is relevant for CMB
    constraints as it will affect the accretion rate of gas onto \bh{}
    in the early Universe \cite{Ricotti_08, bib:AliHaimoud2017a,
      bib:Poulin2017}.  There are two contributions to the variance of
    the PBH velocity distribution: a large-scale piece
    $\langle v_{\rm L}^2 \rangle $, due to linear perturbations on
    scales $k_{\rm box} \lesssim k \lesssim k_{\rm NL}$, and a
    small-scale piece $\langle v_{\rm NL}^2 \rangle$, arising from
    virial motions inside halos.
    
    The large-scale contribution can be estimated from the linearized
    continuity equation for PBHs:
    \begin{align}
      \vec{v}_{\rm L} &= i
                        \frac{\hat{k}}{k} \dot{\delta}_{\bh} = i \hat{k} \frac{a H}{k}
                        \frac{d \ln \delta_\bh}{d \ln a} \delta_\bh.  
    \end{align}
    Since we restrict ourselves to linear scales, the variance of PBH
    density fluctuations per $\ln k$ scales as $k^3$, like their
    initial Poisson distribution. As a consequence the variance of
    $\vec{v}_{\rm L}$ per $\ln k$ scales as $k$, and is dominated by
    the smallest relevant scale, i.e.~the non-linear scale
    $k_{\rm NL}$. By definition, this is the scale at which
    $\delta_\bh \sim 1$. Hence we find
    \begin{align}
      \langle v_{\rm L}^2 \rangle^{1/2} \sim \frac{a H}{k_{\rm NL}}
      \frac{d \ln \delta_\bh}{d \ln a}.  
    \end{align}
    To estimate $k_{\rm NL}$ and $d \ln \delta_\bh/ d \ln a$, we use
    Eq.~\eqref{eq:delta_pbh(a)}, in the limit where PBHs dominate the
    initial density perturbations. We moreover simplify matters by
    taking the limit where baryons cluster like CDM, so that:
    \begin{align}
      \delta_\bh \approx \left(1 +\frac32
      \frac{a}{a_{\rm eq}} f_\bh \right)\delta_\bh^I.  
    \end{align}
    This implies
    \begin{align}
      \frac{d \ln \delta_{\bh}}{d \ln a} &\approx \frac{\frac32 \frac{a}{a_{\rm eq}} f_\bh}{1  + \frac32 \frac{a}{a_{\rm eq}} f_\bh}, \\
      k_{\rm NL}&\approx \left(\frac{2 \pi^2 \overline{n}_{\bh}}{(1 +
                  \frac32 \frac{a}{a_{\rm eq}} f_\bh)^2} \right)^{1/3}, 
    \end{align}
    where $\overline{n}_\bh$ is the comoving number density of PBHs,
    which sets the amplitude of initial Poisson
    perturbations. Numerically, we obtain
    \begin{align} v_{\rm L} \sim 1 \ \textrm{km/s}\ \frac{ f_\bh^{2/3}
        (a/a_{\rm eq})^{1/2}}{\left(1 + \frac32 \frac{a}{a_{\rm eq}}
          f_\bh \right)^{1/3}} .
    \end{align}
    For instance, at $z = 300$, and for $f_\bh = 0.1$, this estimate
    gives $v_{\rm L} \sim 0.5$ km/s. We find that isolated PBHs (which
    do not belong to halos with multiple PBHs) indeed have a
    characteristic velocity of that order, see
    Fig.~\ref{fig:velocity}.

    We now estimate the non-linear velocity, based on the halo model
    developed in the preceding section.  The behaviour of a \bh{} in a
    halo depends significantly on $N_{\bh/\hl}$.  An isolated \bh{}
    will simply sit in the halo center of mass (having seeded its
    formation in the first place), and therefore moves with the halo
    motion.  On the other hand, if the halo hosts a multitude of
    virialized \bh, their motions will include a virial velocity:
    \begin{align}
      \label{eq:Vhl2}
      V_\hl^2 \sim \frac{G M_\hl}{a R_\hl}
    \end{align}
    where we use Eq.~\ref{eq:halomass} for $M_\hl$ and $R_\hl$ is the
    comoving halo radius.  We can estimate $R_\hl$ via:
    \begin{align}
      \Delta_{\rm vir} \bar{\rho}_c = \frac{M_\hl}{\frac{4}{3}\pi (R_\hl)^3}
    \end{align}
    Note that these virial velocities do not depend particularly
    sensitively on the definition of the halo (i.e.~the value of
    $\Delta_{\rm vir}$).  We approximate the probability distribution
    of a $\bh$ in a halo with $N_{\bh/\hl}$ as a 3-dimensional
    Gaussian with variance $\sigma_v^2$:
    \begin{align}
      \frac{dP}{dv}(N_{\bh/\hl}) =
      \sqrt{\frac{2}{\pi}}\frac{v^2}{\sigma_v^3} \exp\left[-\frac{v^2}{2 \sigma_v^2}\right]
    \end{align}
    where $3\sigma_v^2 = V_{\rm HL}^2 + v_{\rm L}^2$ depends on
    $N_{\bh/\hl}$.  The differential distribution of \bh{} velocities
    is obtained by summing over all halos:
    \begin{align}
      \frac{dN_\bh}{dv} = \sum\limits_{N=2}^{\infty} N
      N_\hl(N) \frac{dP}{dv}(N).
    \end{align}
    We compute this distribution using Eq.~\eqref{eq:halomass} for
    $M_\hl$ and Eq.~\eqref{eq:Epstein} for the mass distribution.  Our
    results are shown in Fig.~\ref{fig:velocity} where we see that the
    halo model prediction does a reasonable job in modeling the true
    distribution.  Since Eq.~\ref{eq:Vhl2} is only accurate to factors
    of order unity, this model only provides an order of magnitude
    estimate even if the agreement for $f_\bh=10^{-1}$ appears rather
    good.

    \begin{figure}
      \includegraphics[width=0.45\textwidth]{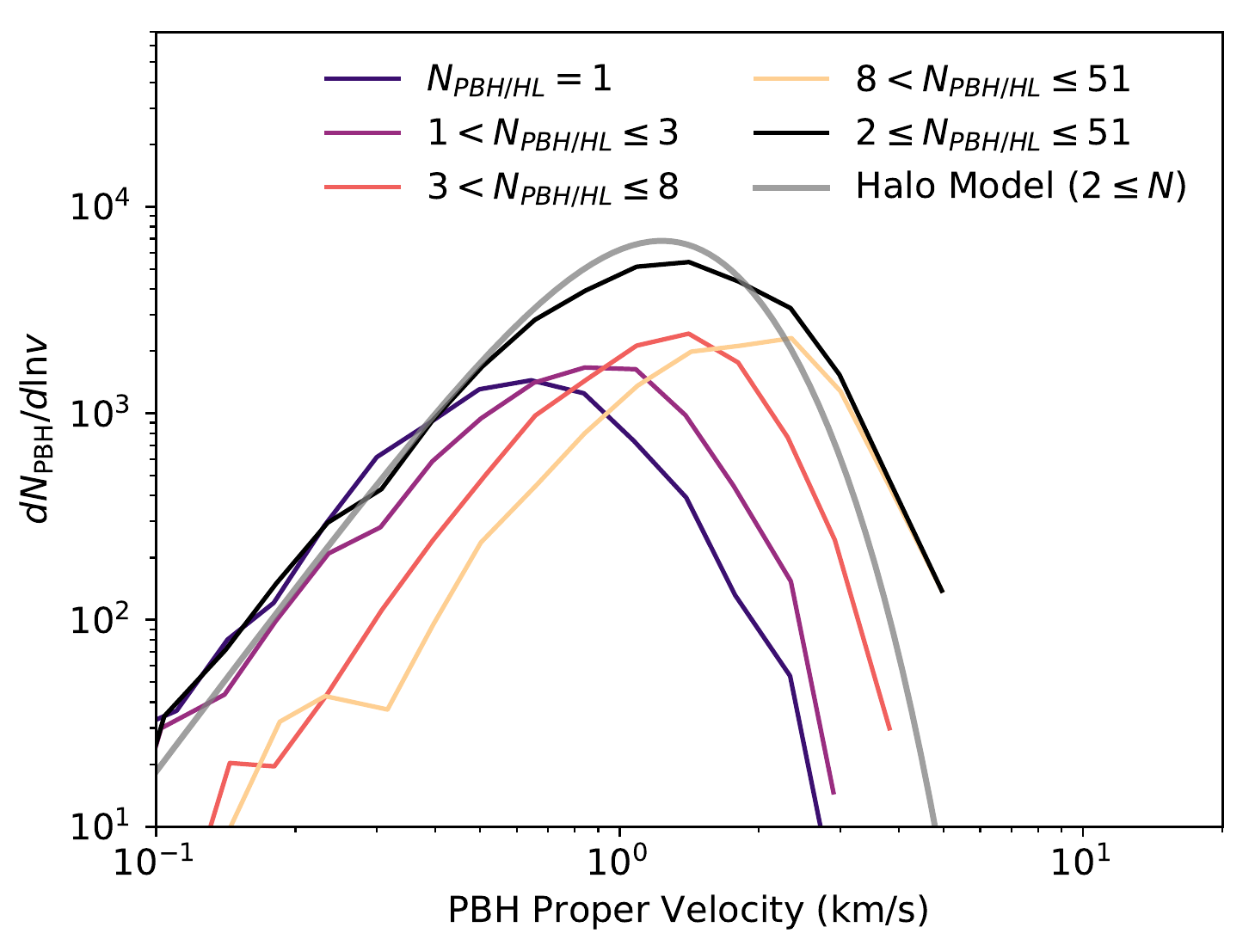}
      \includegraphics[width=0.45\textwidth]{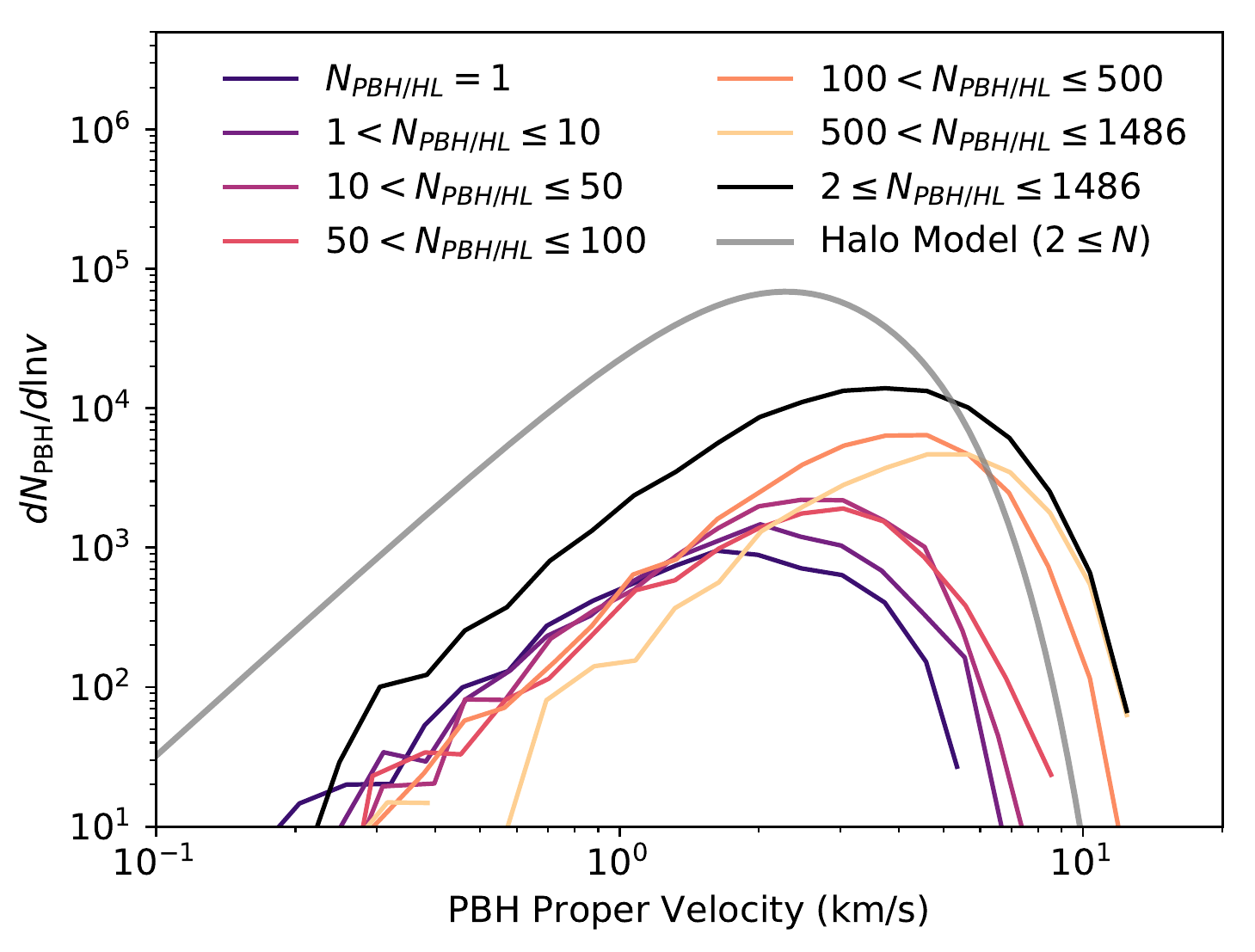}
      \caption{The number of \bh{} per unit velocity as a function of
        \bh{} velocities at $z=99$ for $f_\bh=10^{-1}$ (top) and
        $10^0$ (bottom).  The grey curve shows a prediction via the
        halo model and should be compared to the black curve.  The
        warm coloured lines show components originating from \hl{}
        with different numbers of \bh{}.  Note that we only consider
        \bh{} found by the halofinder in this figure.}
      \label{fig:velocity}
    \end{figure}

    We show the velocity dispersion
    $\sqrt{\langle v^2\rangle - \langle \vec{v} \rangle^2}$ as a
    function of redshift in Fig.~\ref{fig:vz}. We find that it
    increases by an order of magnitude between $f_\bh=10^{-3}$ and
    $f_\bh=10^0$. We compare this velocity dispersion to the
    characteristic large-scale velocity of the gas relative to dark
    matter and the gas sound speed \cite{bib:Tseliakhovich2010}
    \begin{align}
      v_{\rm rel} &\approx 30~ \kms \frac{1+z}{1000},\\
      c_s &\approx 6~\kms \sqrt{\frac{1+z}{1000}}.  
    \end{align}
    We see in Fig.~\ref{fig:vz} that the typical non-linear motions
    are smaller than the baryon sound speed for $z \gtrsim 300$, even
    for $f_\bh \rightarrow 1$. This suggests that, even in patches of
    low relative velocity, baryons may not be efficiently captured in
    the first PBH + PDM halos until $z \lesssim 300$. In regions with
    typical relative velocities, baryons may not be efficiently
    accreted until $z \lesssim 100$.  Thus, even though small-scale
    dark matter structures form much earlier than in standard PDM
    cosmology, it is unclear whether and when they would acquire a
    significant baryon content. It would be interesting to address
    this question with dedicated hydrodynamical simulations.
    
    Non-linear velocities could in principle affect CMB limits to
    accreting PBHs \cite{Ricotti_08, bib:AliHaimoud2017a,
      bib:Poulin2017}. At equal gas density, larger velocities would
    decrease the accretion rate. However, once the gas gets bound to
    halos, it may get significantly denser than the cosmological
    average. Within the most conservative ``collisional ionization"
    limit of \cite{bib:AliHaimoud2017a}, we have checked that
    completely switching off black hole accretion at $z < 300$ has a
    negligible impact on CMB bounds. Conversely, increasing the PBH
    luminosity by a factor of 10 at $z < 300$ tightens CMB upper
    limits by no more than $\sim 10\%$, and only at the highest mass
    end. It is therefore unlikely that CMB limits are significantly
    affected by non-linear velocities, though a definitive conclusion
    would require more detailed modeling of both the baryon
    distribution, as well as the accretion process.  Note that
    Ref.~\cite{Hutsi_19} recently reached the same conclusion, using
    analytic estimates of the PBH velocity distribution, although they
    did not explicitly check the effect on CMB limits as we do here.
    
    \begin{figure}
      \includegraphics[width=0.45\textwidth]{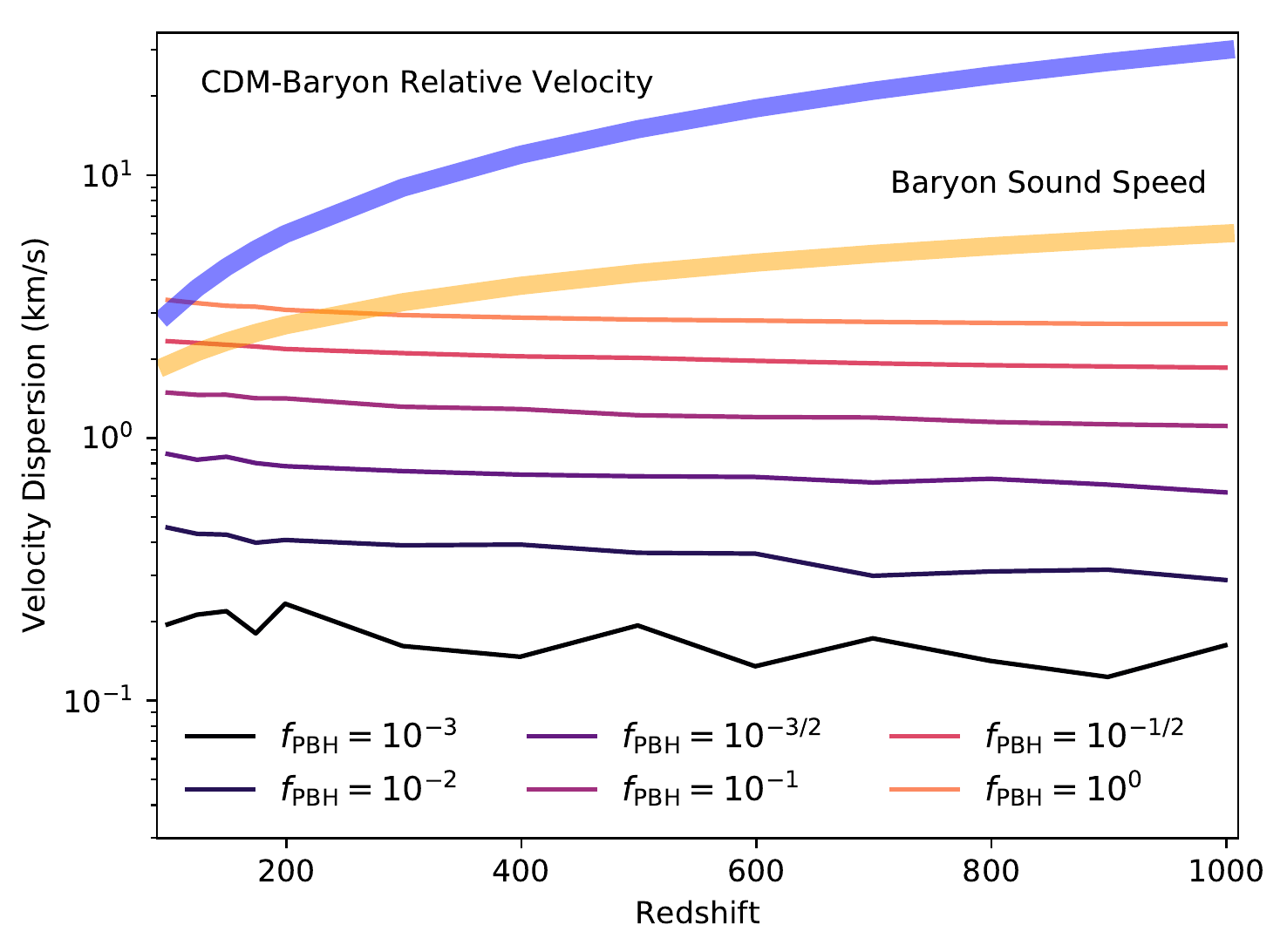}
      \caption{The \bh{} velocity dispersion as a function of $f_\bh$
        and redshift.  \bh{} clustering increases the velocities of
        \bh{}, but not to speeds comparable to the relative velocity
        effect.}
      \label{fig:vz}
    \end{figure}

  \end{subsection}

  \begin{subsection}{Approximate self-similarity}
    While our results have specifically focused on the case where
    $M_\bh=20M_\odot h^{-1}$, we note that our simulations are
    approximately self-similar, hence our results can be easily
    transposed to different PBH masses. First of all, Hamilton's
    equations are invariant under the following rescaling of
    positions, velocities and masses:
    \begin{align}
      x\rightarrow\lambda x\nonumber\\
      v\rightarrow\lambda v\nonumber\\
      M\rightarrow\lambda^3 M \label{eq:scaling}
    \end{align}
    Secondly, the initial random distribution of PBHs is exactly
    scale-invariant. Finally, the adiabatic initial perturbations are
    not strictly scale invariant, but are nearly so on the scales of
    interest.
        
    Our simulation setup explicitly respects the self-similarity of
    the underlying equations. Before applying displacement fields, the
    \dm{} particles are set on a cubical lattice, a structure which is
    clearly independent of scale. Similarly, the distribution of \bh{}
    particles is completely random, and therefore also cannot depend
    on box size.  The simulation box size only enters explicitly when
    setting the adiabatic initial conditions. Subsequent gravitational
    evolution only depends on the particles' numerical mass
    (Eq.~\ref{eq:simmass}) which is again independent of any physical
    scale. Our simulation setup therefore explicitly preserves the
    approximate self-similarity given by Eq.~\eqref{eq:scaling}.
    
    Of course, approximate self-similarity does not extend to
    arbitrary scales. In particular, for $M_\bh\gtrsim 10^4 M_\odot$,
    the box size exceeds the horizon scale at the starting redshift.
    This is unlikely to be a severe issue, however, since there is not
    much gravitational evolution until matter-radiation equality, at
    which point the horizon size exceeds the box size as long as
    $M_\bh \lesssim 10^{11} M_{\odot}$. More importantly, while the
    simulation scales satisfy $k \propto M_\bh^{-1/3}$, the scale
    $k_*$ of the enhancement in the primordial power spectrum required
    to form PBHs has a steeper scaling $k_* \propto M_\bh^{-1/2}$
    \cite{Sasaki_18}. Given our fiducial box and grid size, the grid
    scale would exceed $2 \pi/k_*$ for $M_\bh \gtrsim 10^7 M_\odot$,
    and we could no longer extrapolate the primordial large-scale
    power spectrum all the way down to the grid scale. On the other
    hand, there does not appear to be an obvious limitation to
    rescaling our simulations to smaller PBH masses.
    
  \end{subsection}

\end{section}

\begin{section}{Discussion and Conclusion}
  We have studied how structure forms in \lpbh{} cosmologies
  containing a mixture of primordial black holes (\bh) and standard
  particle dark matter (\dm).  Our results depend sensitively on what
  fraction of the CDM is in the form of PBH.  For
  $f_\bh\lesssim z\times10^{-4}$, the \bh{} are generally isolated
  from one another and accrete \dm{} to form halos.  For
  $f_\bh\gtrsim z\times10^{-4}$, there is significantly more
  clustering of \bh{} to form much larger halos. We find that halo
  masses are nearly linearly proportional to the number of \bh{} they
  contain and that the halo mass function is well described via
  Poisson statistics.  Isolated halos containing only a single black
  hole tend to form steep power law \dm{} distributions, regardless of
  $f_\bh$.  On the other hand, such steep profiles do not occur when
  the halos contain many virialized \bh.  We quantified the nonlinear
  velocities of \bh{} and find them strongly subdominant to the
  relative velocity between gas and CDM for $z\gtrsim 300$. We showed
  that they should not significantly affect CMB constraints to
  accreting PBH \citep{bib:AliHaimoud2017a}.  While we only ran
  simulations with $M_\bh=20 ~h^{-1} M_\odot$, we argued that our
  simulations are approximately self-similar, and can be simply
  rescaled to arbitrary PBH mass, up to a maximum mass of
  $\sim 10^7 M_{\odot}$.
  
  Current limitations on our simulations are largely computational.
  For instance, it would be beneficial for the $f_\bh=1$ simulation to
  evolve a larger volume so that the adiabatic perturbations are
  resolved and box-scale modes do not become nonlinear.  Increasing
  the number of \dm{} particles would also lead to better resolved
  halo inner structures.  It is also challenging to extrapolate our
  results to the present day which impacts our ability to determine if
  these \bh{} minihalos survive and may be present in the galactic
  halo.  This is of crucial importance for some constraints due to
  \dm{} annihilation \citep{bib:Lacki2010} or astrometry
  \cite{bib:VanTilburg2018}.

  With the development of the numerical simulation code, computations
  that are challenging analytically become more tractable. We are
  particularly interested in studying the Poisson clustering of \bh{},
  which should lead to the formation of many binaries deep in the
  radiation era.  If such binaries survive until the present, then
  {\sc LIGO} will have already put significant constraints on $f_\bh$
  \citep{bib:AliHaimoud2017b}.  However, these constraints rely on
  many approximations that require numerical verification.
  Ref.~\citep{bib:Kavanagh2018} studied the interaction of pairs of
  \bh{} initially clothed by \dm{}.  They find that as the \bh{}
  orbit, they lose their \dm{} halos and dynamical friction effects
  tend to circularize and shrink the orbit.  Overall, however, the
  effects are modest provided the initial eccentricity is large.  A
  different mechanism for disrupting the binaries is nontrivial tidal
  forces.  Ref.~\citep{bib:Raidal2018} simulated binary \bh{}
  surrounded by other \bh{} (but not \dm{}).  They find that binaries
  are frequently disrupted for $f_\bh \gtrsim 10^{-1}$.  While our
  numerical setup does not have high enough spatial resolution to
  resolve individual binary orbits, they should allow us to understand
  the tidal field the binary is immersed in.  We expect to obtain
  complementary information to
  Refs.~\citep{bib:Kavanagh2018,bib:Raidal2018} since our simulations
  contain both \dm{} and \bh{}, can go an order of magnitude lower in
  redshift, and probe halos with many more \bh.

  Another interesting computation is the effect of \lpbh{} on cosmic
  dawn. In the standard picture, in the absence of PBHs, first stars
  are expected to typically form when CDM halos reach a mass of
  $10^6 M_\odot$ \citep{bib:Visbal2012}. Given that halos tend to form
  much earlier in $\Lambda$PBH (even with a subdominant fraction of DM
  in PBHs), it is possible that the first lights in the Universe turn
  on earlier.  Such a modification to early star formation by \bh{}
  has been studied analytically in Ref.~\citep{bib:Kashlinsky2016} who
  found it could explain the remnant cosmic infrared background
  fluctuations that cannot be explained via observed galaxies.  We
  expect that this analysis can be improved by extracting the virial
  temperature, $T\propto M/R$, as a function of halo mass since the
  formation of first stars is largely sensitive to the gas
  temperature.  It may become necessary to simulate the baryons as
  well, although this would require modelling their CMB interactions,
  and later on radiative processes from star formation.

  Another area to explore is how much \bh{} clustering is affected
  when the adiabatic power spectrum is not the standard \lcdm{} one
  but rather is enhanced on certain scales. Such a change in the
  adiabatic power spectrum is also likely to affect the \bh{} binaries
  \citep{bib:Garriga2019} and cosmic dawn \citep{bib:Hirano2015} as
  well.  We intend to tackle these interesting questions in future
  publications.

\end{section}

\begin{section}{Acknowledgements}

  We acknowledge valuable technical support and discussions with
  Hao-ran Yu, Ue-Li Pen and JD Emberson. We thank Daniel Grin, Neal
  Dalal, Ravi Sheth and Joe Silk for useful conversations.  This work
  was made possible in part through the NYU IT High Performance
  Computing resources, services, and staff expertise. We acknowledge
  the use of {\sc NumPy} \citep{bib:Walt2011}, {\sc SciPy}
  \cite{bib:Jones2001}, {\sc Matplotlib} \citep{bib:Hunter2007} and
  NASA's Astrophysics Data System Bibliographic Services.  This
  research is supported by the National Science Foundation under Grant
  No.~1820861.

\end{section}

\appendix

\begin{section}{Spherical top-hat collapse in matter and radiation
    era}
  \label{app:spherical-collapse}

  The famous value of the linearly-extrapolated density at collapse,
  $\delta_{\rm cr} = 1.686$, is only valid for collapse that occurs
  entirely in matter domination. This is an accurate approximation for
  halos collapsing around redshift 0 (or rather, around redshift of a
  few, before dark-energy domination). It needs not be accurate for
  halos collapsing at $z \gtrsim 100$, of interest here, for which
  radiation contributed a significant part of the energy density, at
  least in the initial phases of the evolution. Furthermore, prior to
  kinematic decoupling around $z \sim 10^3$, baryons are subject to
  strong Compton drag, which prevents their clustering on all
  scales. On scales smaller than the baryon Jeans scale, they remain
  unclustered even after decoupling from CMB photons. To account for
  these effects, we generalize the spherical top-hat collapse model
  \cite{bib:Gunn1972} to a halo of cold dark matter collapsing in a
  Universe comprised of matter and radiation (the latter assumed fully
  homogeneous), assuming only the cold dark matter clusters, and
  baryons remains homogeneous. This setup is similar in spirit to that
  of Refs.~\cite{bib:LoVerde2014}, who studied spherical top-hat
  collapse in the presence of free-streaming neutrinos. See also
  Ref.~\cite{bib:Naoz2007} for a similar study, with different setup
  (we assume baryons remain unclustered throughout the collapse, being
  interested in sub-Jeans scale perturbations).  We restrict ourselves
  to deeply sub-horizon scales. We define
  $\gamma \equiv \Omega_c/\Omega_m$ to be the fraction of clustered
  matter.

  While the density perturbations become non-linear, metric
  perturbations remain small, and we may assume a perturbed FLRW
  metric. We assume a uniform, top-hat dark matter overdensity
  $\delta(a)$, within a comoving radius $R(a)$. The constant CDM mass
  contained in this region is
  $M = \frac{4 \pi}{3} \overline{\rho}_{c, 0} (1 + \delta) R^3 =
  \frac{4 \pi}{3} \overline{\rho}_{c, 0} (1 + \delta_i) R_i^3$,
  where $\overline{\rho}_{c, 0} = \gamma \rho_{m, 0}$ is the
  background comoving CDM density, and $R_i$ and $\delta_i$ are the
  initial radius and overdensity, respectively. Combining the geodesic
  equation for $R(a)$ with the Poisson equation, and defining
  $\chi \equiv R/R_i$, we obtain the following differential equation
  for $\chi(s = a/a_{\rm eq})$:
  \begin{align}
    2s(1+ s) \chi''(s) + (2 + 3 s)\chi'(s)
    = \gamma\left(\chi - \frac{1+ \delta_i}{\chi^2}\right).  
  \end{align}
  This equation must be solved subject to initial conditions
  $\chi(0) = 1, \chi'(0) = - \frac{\gamma}2 \delta_i$, the latter
  being the only well-behaved initial condition for the initial
  velocity. Given an initial overdensity $\delta_i$, we solve this
  equation up until the time $s_{\rm coll}$ at which $\chi$ vanishes.

  The CDM overdensity is related to $\chi$ through
  $(1 + \delta) = (1 + \delta_i)/\chi^3$. We define
  $\delta_{\rm lin}(s)$ to be the solution of the \emph{linearized}
  fluid equation for the overdensity. Assuming an initially constant
  overdensity, we have $\delta_{\rm lin}(a) = D_+(a) \delta_i$, where
  $D_+$ is the linear growth rate computed in Section
  \ref{sec:pert-theory}. For a given $\delta_i$, we compute the
  linearly-extrapolated overdensity at the time of collapse
  $s_{\rm coll}$,
  $\delta_{\rm lin}(s_{\rm coll}) =D_+(s_{\rm coll}) \delta_i$. We
  show this critical density as a function of collapse scale factor in
  Fig.~\ref{fig:delta_crit}.
  \begin{figure}
    \includegraphics[width = \columnwidth]{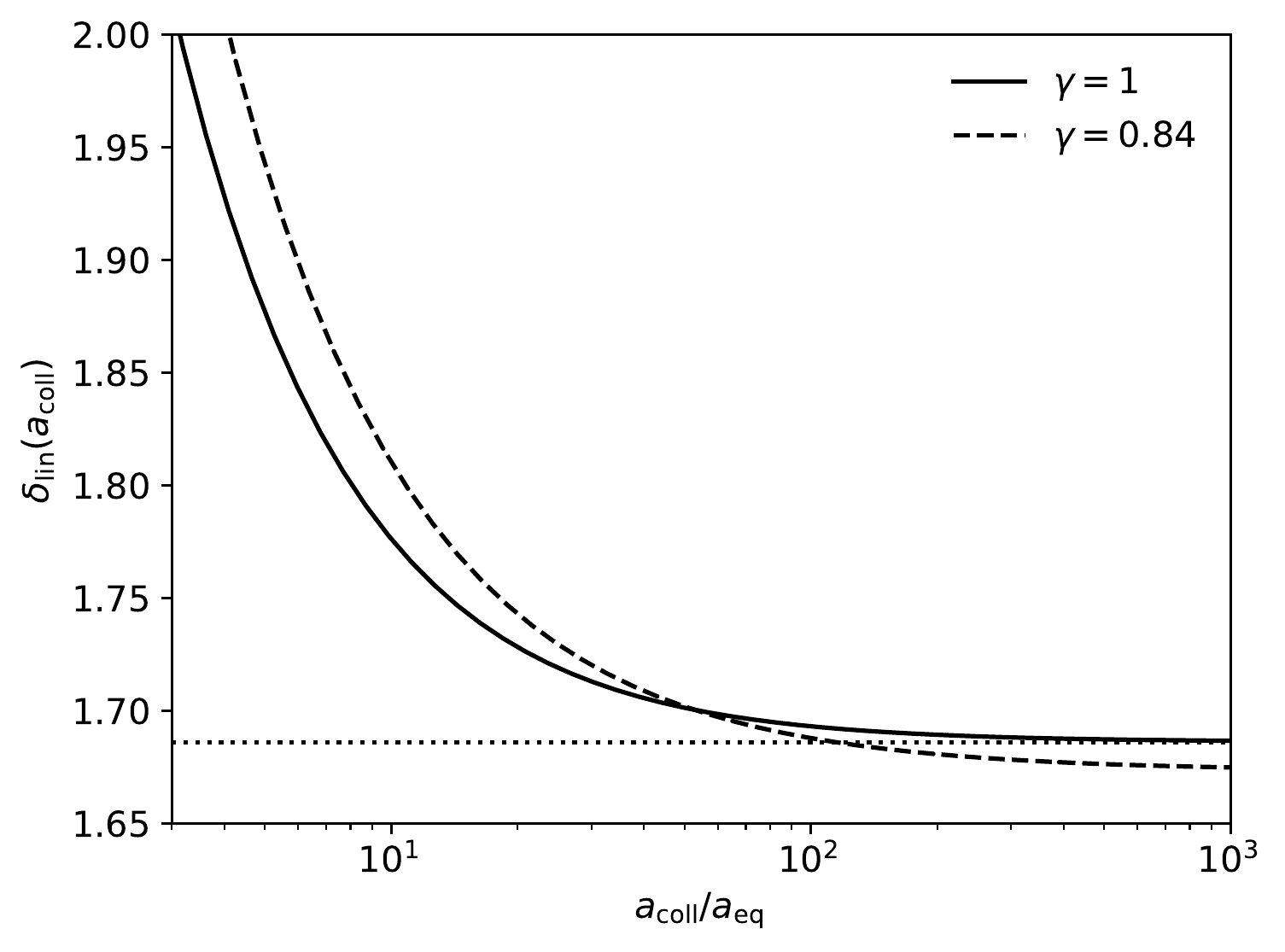}
    \caption{Linearly-extrapolated dark matter overdensity at collapse, as
      a function of collapse scale factor, assuming a homogeneous
      radiation background. The dotted line shows the standard value
      $\delta_{\rm cr} = 1.686$, which holds when collapse occurs deep
      in the matter-domination era. The solid line shows the critical
      overdensity when all the matter clusters. The dashed line show the
      corresponding value when only a fraction $0.84$ of the matter
      clusters.}
    \label{fig:delta_crit}
  \end{figure}

\end{section}

\begin{section}{Convergence Tests}
  \label{app:convergence}

    In this Appendix we seek to check the dependence of our results on
    three different factors: (1) the force softening length, (2) the
    initial conditions, and (3) the number of \dm{} particles.  To do
    this we setup simulations with $10^4$ \bh{} and $2\times128^3$
    \dm{} particles each with {\it no} initial perturbations (i.e.~a pure
    Poissonian distribution of \bh{} and lattice distribution of \dm).
    We run four of these simulations with varying softening length
    $r_{\rm soft} = 0.2,0.1,0.05$ and $0.025$ fine cells.  These
    simulations can then be compared to the reference $f_\bh=10^{-1}$
    simulation in the main text.  Note that since the particle number
    has changed, it is the $0.05$ softening length that is equivalent
    to the reference simulation.  The results we expect to be most
    sensitive to softening are the profiles of halos with many \bh{}
    and the \bh{} virial velocities.  We show the former in
    Fig.~\ref{fig:convergence_profile}.  While it is clear that
    softening plays a significant role on the inner parts of the halo,
    it appears that smaller softening lengths produce even flatter cores.  Nonetheless, higher
    particle numbers are still needed to confirm.  The \bh{}
    velocity dispersion as a function of redshift are shown in
    Fig.~\ref{fig:convergence_vz}.  We find that our choice of
    softening length actually yields the largest dispersion.  We
    expect this to not change our conclusions as we had already argued
    that these velocities are small.

  We have also considered and tested using different softening length
  for the \bh{} and the \dm{}, since ideally the \bh{} would be
  softened significantly less.  This is not an issue for simulations
  with low number density where \bh{} never are nearby one another,
  but is an issue at higher number densities.  For simulations where
  \bh{} are numerous enough to require softening in the \bh-\bh{}
  force, we find that the smallest softening
  length determines total runtime.  There is therefore no benefit to
  not using the smallest softening length for all particles.  This
  likely would not be true if different particles could have different
  timesteps, in which case it could be beneficial to set the \bh{}
  softening length lower.  However, such a feature is currently not
  implemented in \cpm.

  \begin{figure}
    \includegraphics[width=0.45\textwidth]{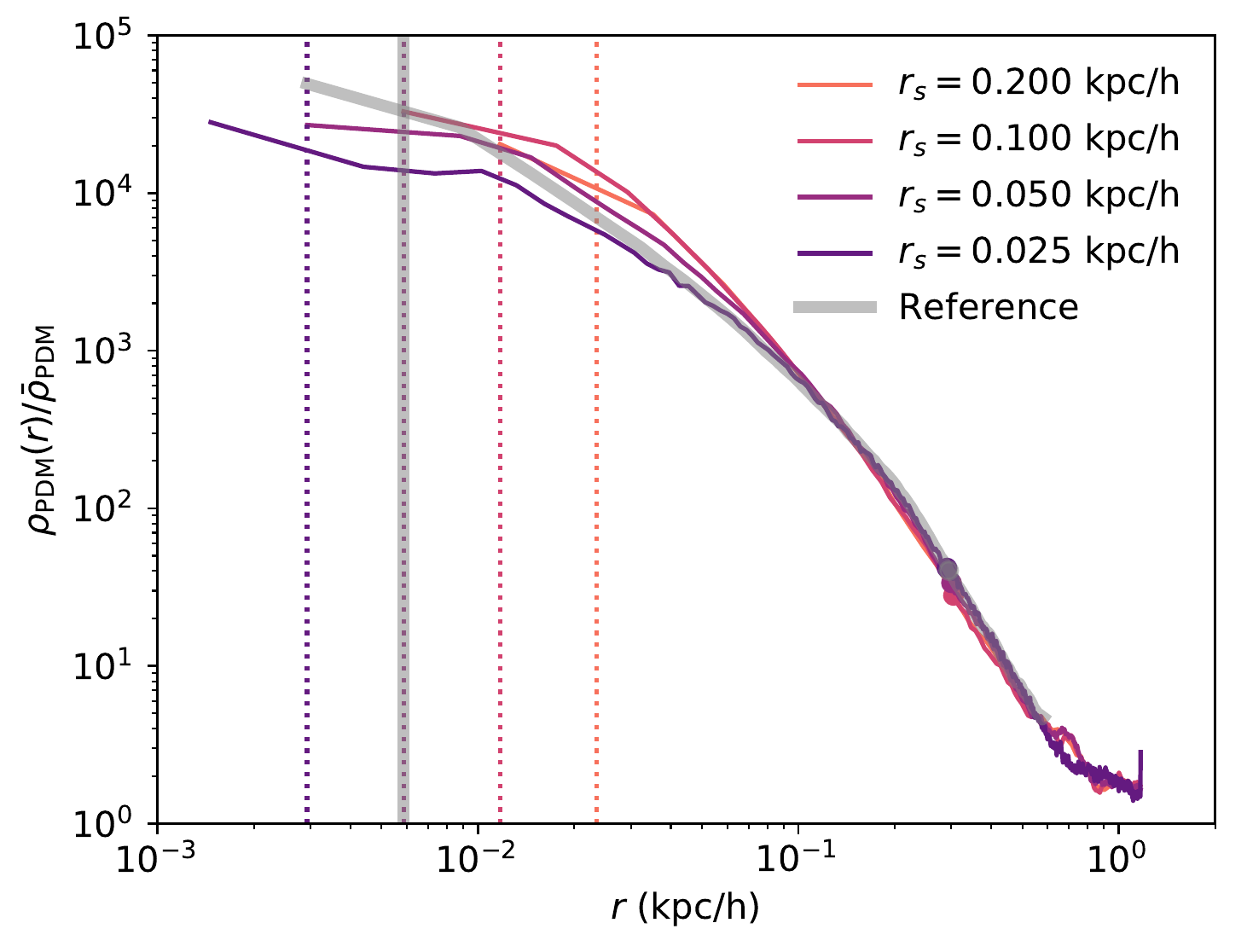}
    \caption{Convergence test of Fig.~\ref{fig:profile_13_17}. The
      grey band shows the result presented in the main text, whereas
      the other curves arise from simulations with 1/8 the number of
      \dm{} particles, no initial conditions, and varying the
      softening length.}
    \label{fig:convergence_profile}
  \end{figure}

  \begin{figure}
    \includegraphics[width=0.9\columnwidth]{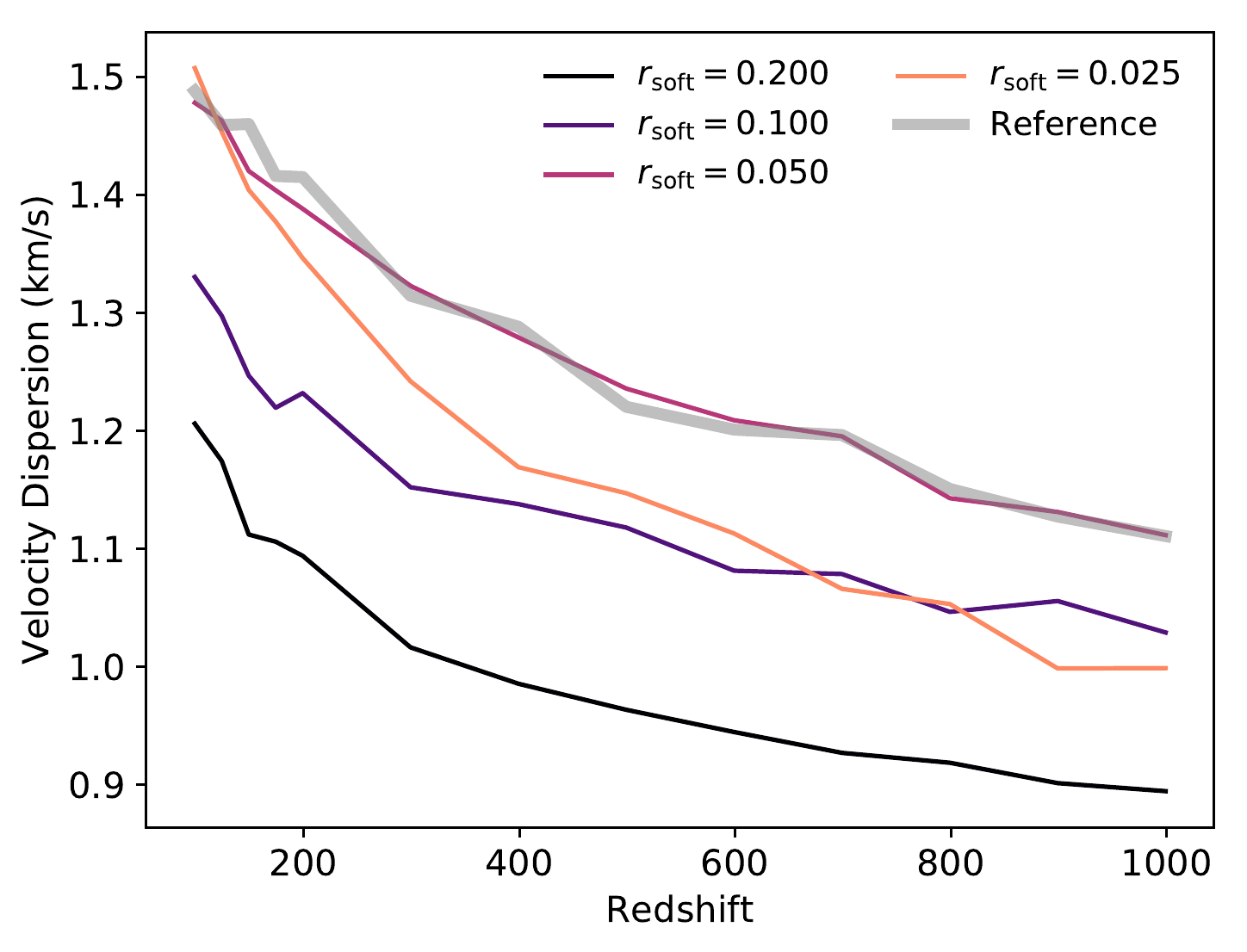}
    \caption{Convergence test of Fig.~\ref{fig:vz}.  Otherwise same description
      as Fig.~\ref{fig:convergence_profile}.}
    \label{fig:convergence_vz}
  \end{figure}

\end{section}

\newpage

\bibliographystyle{apsrev} \bibliography{thebib}

\end{document}